\documentclass[twocolumn]{aa}
\usepackage{graphicx}
\usepackage{txfonts}
\usepackage{rotating}
\usepackage{aalongtable,lscape}
\def\XMM{{\sl XMM-Newton}}
\def\ASCA{{\sl ASCA}}
\def\ROSAT{{\sl ROSAT}}
\def\Chan{{\sl Chandra}}
\def\cgs{erg cm$^{-2}$ s$^{-1}$}

\begin{document}
   \title{The XMM-Newton serendipitous survey}

   \subtitle{VI. The X-ray Luminosity Function}

   \author{J. Ebrero\inst{1,2}
          \and
          F. J. Carrera\inst{1}
          \and
          M. J. Page\inst{3}
          \and
          J. D. Silverman\inst{4}
          \and
          X. Barcons\inst{1}
          \and
          M. T. Ceballos\inst{1}
          \and
          A. Corral\inst{1,5}
          \and
          R. Della Ceca\inst{5}
          \and
          M. G. Watson\inst{2}
          }

   \offprints{J. Ebrero, \email{ecarrero@ifca.unican.es, jec33@star.le.ac.uk}}

   \institute{Instituto de F\'\i sica de Cantabria (CSIC-UC), Avenida de los
     Castros, 39005 Santander, Spain
     \and
     Department of Physics and Astronomy, University of Leicester, University Road, LE1 7RH, Leicester, UK
     \and
     Mullard Space Science Laboratory, University College London, Holmbury St. Mary, Dorking, Surrey, RH5 6NT, UK
     \and
     Institute of Astronomy, Swiss Federal Institute of Technology (ETH H\"onggerberg), CH-8093 Z\"urich, Switzerland
     \and
     INAF-Osservatorio Astronomico di Brera, Via Brera 28, 20121 Milano, Italy
             }

   \date{Received <date> / Accepted <date>}

   \abstract
 {}
 {To study the cosmological evolution of Active Galactic Nuclei (AGN) is one of the main goals of X-ray surveys. To accurately determine the intrinsic (before absorption) X-ray luminosity function, it is essential to constrain the evolutionary properties of AGN and therefore the history of the formation of supermassive black holes with cosmic time.}
{In this paper we investigate the X-ray luminosity function of absorbed ($\log N_H > 22$) and unabsorbed AGN in three energy bands (Soft: 0.5-2~keV, Hard: 2-10~keV and Ultrahard: 4.5-7.5~keV). For the Hard and Ultrahard sources we have also studied the $N_H$ function and the dependence of the fraction of absorbed AGN with luminosity and redshift. This investigation is carried out using the XMS survey along with other highly complete flux-limited deeper and shallower surveys in all three bands for a total of 1009, 435 and 119 sources in the Soft, Hard and Ultrahard bands, respectively. We have modelled the instrinsic absorption of the Hard and Ultrahard sources ($N_H$ function) and computed the X-ray luminosity function in all bands using two methods. The first makes use of a modified version of the classic $1/V_a$ technique, while the second performs a Maximum Likelihood fit using an analytic model and all available sources without binning.}
 {We find that the X-ray luminosity function (XLF) is best described by a Luminosity-dependent Density Evolution (LDDE) model. Our results show a good overall agreement with previous results in the Hard band, although with slightly weaker evolution. Our model in the Soft band present slight discrepancies with other works in this band, the shape of our present day XLF being significantly flatter. We find faster evolution in the AGN detected in the Ultrahard band than those in the Hard band.}
 {The results reported here show that the fraction of absorbed AGN in the Hard and Ultrahard  bands is dependent on the X-ray luminosity. We find evidence of evolution of this fraction with redshift in the Hard band whereas in the Ultrahard band there is none, possibly due to the low statistics. Our best-fit XLF shows that the high-luminosity AGN, detected in all bands, exhibit a similar behaviours and are fully formed earlier than the less luminous AGN. The latter sources account for the vast majority of the accretion rate and mass density of the Universe, according to an anti-hierarchical black hole growth scenario.}


   \authorrunning{J. Ebrero et al.}
   \titlerunning{The \XMM{} serendipitous survey. VI. The X-ray luminosity function.}
   
   \maketitle
%

\section{Introduction}

One of the main goals of X-ray surveys is to study the cosmological properties of Active Galactic Nuclei (AGN) as they are strongly linked to the accretion history of the Universe and the formation and growth of the supermassive black holes that are believed to reside in the centre of all galaxies, active or not (Kormendy \& Richstone \cite{Kormendy95}, Magorrian et al. \cite{Magorrian98}, Richstone et al. \cite{Richstone98}). The first studies in this field were constrained to the soft X-ray band ($\leq$2~keV) which could be biased against absorbed AGN (Maccacaro et al. \cite{Maccacaro91}, Boyle et al. \cite{Boyle93}, Page et al. \cite{Page97}, Miyaji et al. \cite{Miyaji00}). Therefore, hard X-ray surveys ($>$2~keV) are essential to describe the luminosity function of the AGN population, including obscured AGN which should be the main contributors to the cosmic X-ray background (see Setti \& Woltjer \cite{Setti89} and Fabian \& Barcons \cite{Fabian92} for a review). Moreover, the large majority of energy density generated by accretion power seems to take place in obscured AGN (Fabian et al. \cite{Fabian98}), as demonstrated by the integrated energy density of the cosmic X-ray background (see e.g. Comastri et al. \cite{Comastri95}, Gilli et al. \cite{Gilli07}). Ignoring the obscured AGN population could therefore bias our understanding of the cosmic evolution of the structures in the X-ray Universe. Furthermore, it has been confirmed that the fraction of absorbed AGN decreases with increasing X-ray luminosity in X-ray selected samples of AGN (Ueda et al. \cite{Ueda03}, Steffen et al. \cite{Steffen03}) as well as in optically selected AGN (i.e. Simpson \cite{Simpson05}). In addtion, some authors claim to have found a positive evolution of the fraction of absorbed AGN with redshift (La Franca et al. \cite{LaFranca05}, Ballantyne et al. \cite{Ballantyne06}, Treister \& Urry \cite{Treister06} and, more recently, Hasinger \cite{Hasinger08}). The evolution of the fraction of the absorbed/obscured AGN is still a subject of debate and its study is therefore one of the primary targets of AGN surveys.

Hard X-ray photons with energies between 2~and 10~keV can pass through large amounts of matter without being absorbed and hence are very useful to detect absorbed sources with intrinsic column densities up to $N_H < 10^{24}$~cm$^{-2}$, although they are not energetic enough to escape from Compton-thick objects ($N_H \gtrsim 10^{24}$~cm$^{-2}$). Previous hard X-ray surveys in the 2-10~keV band (Cagnoni et al. \cite{Cagnoni98}, Fiore et al. \cite{Fiore99}, Giacconi et al. \cite{Giacconi01}, Baldi et al. \cite{Baldi02}, Harrison et al. \cite{Harrison03}, Alexander et al. \cite{Alexander03}) have provided least biased and highly complete samples of sources, many of them being obscured AGN. Notwithstanding this, an important fraction of the AGN detected in deep fields fail to provide good quality X-ray information, and their optical counterparts are far too faint to make reliable spectroscopic identifications of the totality of the samples.

Previous works have studied the X-ray luminosity function (XLF, hereafter) of AGN, mainly in the 0.5-2~keV and 2-10~keV bands. For instance, Miyaji et al. (\cite{Miyaji00}) and Hasinger et al. (\cite{Hasinger05}) studied the 0.5-2~keV XLF of Type-1 AGN testing a variety of models. Their results have ruled out both Pure Luminosity and Pure Density Evolution models in favour of a Luminosity Dependent Density Evolution model which best describes the observed XLF of these sources. In this paper we use a sample that contains $\sim$30\% more AGN than the Miyaji et al. (\cite{Miyaji00}) sample and is comparable to that of Hasinger et al. (\cite{Hasinger05}), including not only Type-1 AGN but also sources identified as Type-2 AGN in order to gain an insight into the evolutionary properties of the whole AGN population at soft X-rays.

Barger et al. (\cite{Barger05}) measured the cosmic evolution of AGN in the \Chan{} Deep Field inthe 2-8~keV energy band finding it consistent with a Pure Luminosity Evolution model. Silverman et al. (\cite{Silverman08}) studied the XLF of high redshift AGN in the 2-8~keV band but they did not consider the intrinsic absorption of the sources due to the limited count statistics. Other works in the 2-10~keV band such as Ueda et al. (\cite{Ueda03}) and La Franca et al. (\cite{LaFranca05}) have taken into account the $N_H$ distribution of the sources when calculating the XLF, but an important fraction of the $N_H$ values were derived from the hardness ratios of the sources rather than from a spectral analysis. This may introduce a certain degree of uncertainty since a template spectrum has to be assumed to compute the $N_H$. The XLF of very hard sources ($>$5~keV) is almost unexplored so far with the exception of the work by Della Ceca et al. (\cite{DellaCeca08}), who studied the de-evolved ($z=0$) luminosity function of absorbed and unabsorbed AGN in the 4.5-7.5~keV band but were unable to perform detailed evolutionary studies of the absorbed population due to the low statistics.

In this work we use the \XMM{} Medium Survey (XMS, Barcons et al. \cite{Barcons07}), along with other highly complete deeper and shallower surveys, to compute the X-ray luminosity function in several energy bands. Furthermore, given the availability of high-quality X-ray spectral information in the XMS, we are able to model the intrinsic absorption of the hardest sources (4.5-7.5~keV) as a function of the X-ray luminosity up to column densities of $\sim 10^{24}$~cm$^{-2}$. These issues are key tools to probe the accretion history of the Universe across cosmic time. Thanks to the extremely high identification completeness of the XMS sample ($\sim$96\% in the 0.5-2~keV band, and $\sim$85\% in the 2-10~keV and 4.5-7.5~keV bands) and the accompanying surveys we have assembled an overall sample of $\sim$1000 identified AGN in the 0.5-2~keV, $\sim$450 identified AGN in the 2-10~keV band, and $\sim$120 identified AGN in the 4.5-7.5~keV bands, leading to one of the largest and most complete sample up to date in all three energy bands.

\begin{figure}
\centering
\hbox{
\includegraphics[width=6.5cm,angle=270.0]{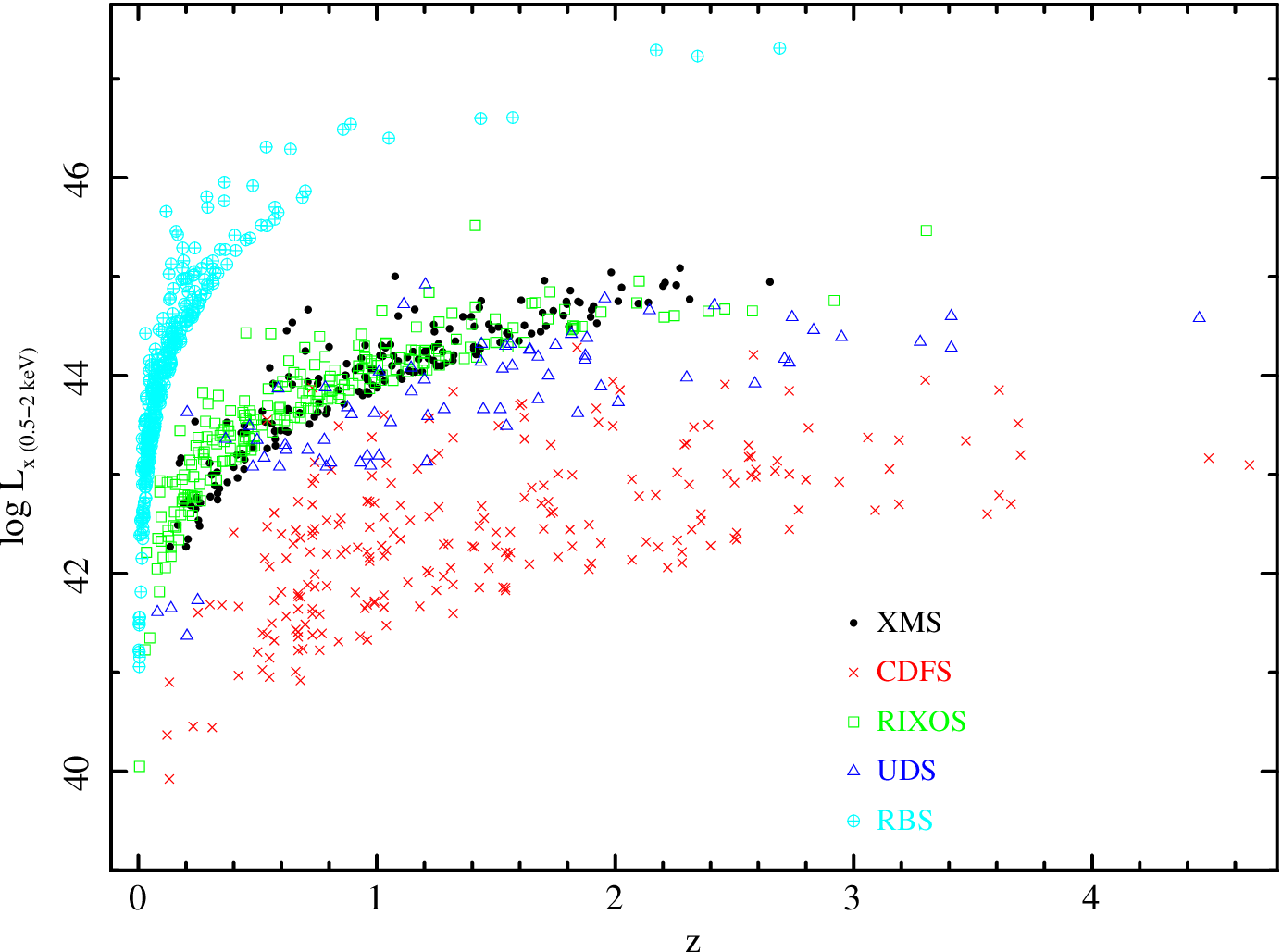}
}
\hbox{
\includegraphics[width=6.5cm,angle=270.0]{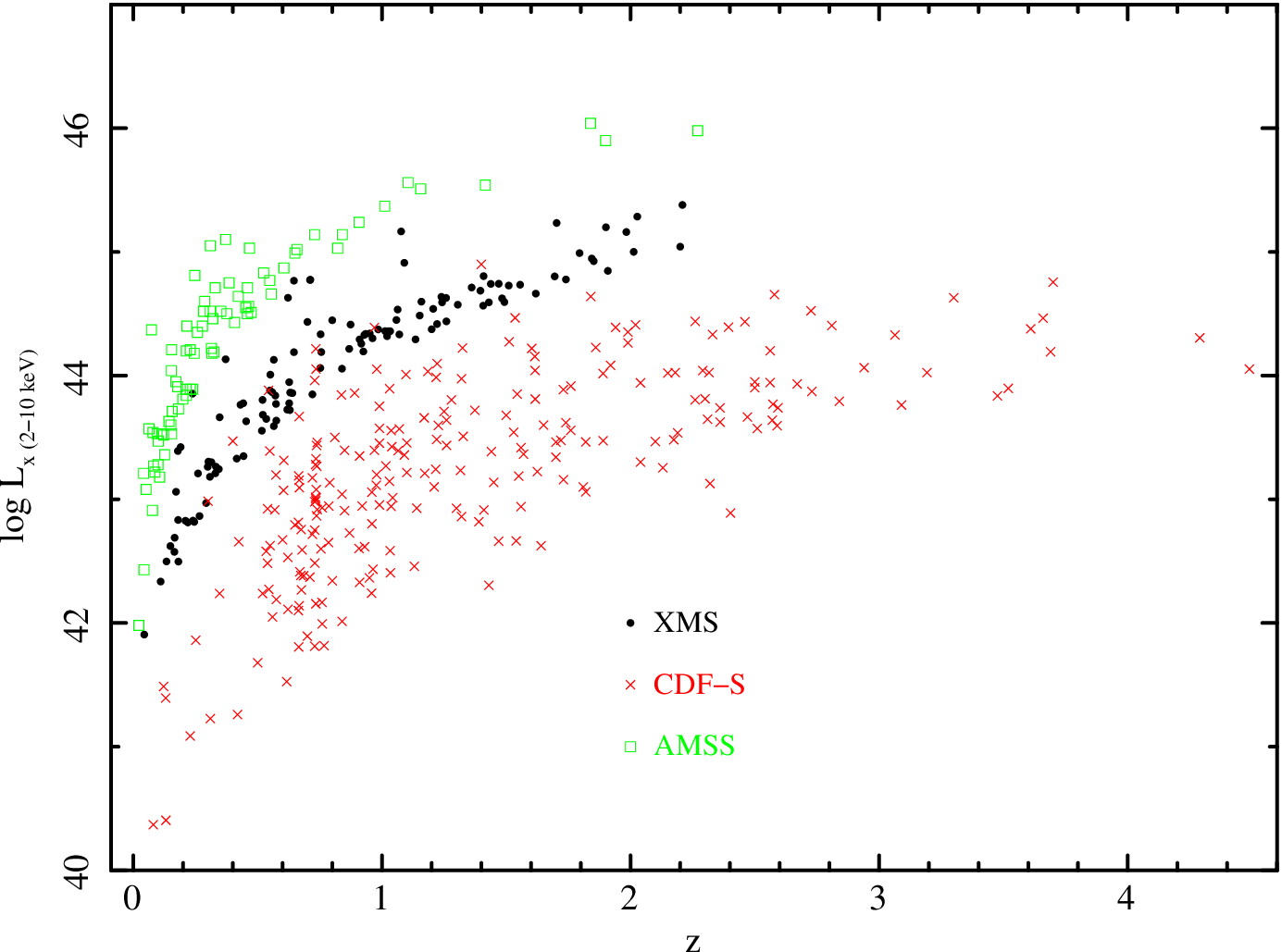}
}
\hbox{
\includegraphics[width=6.5cm,angle=270.0]{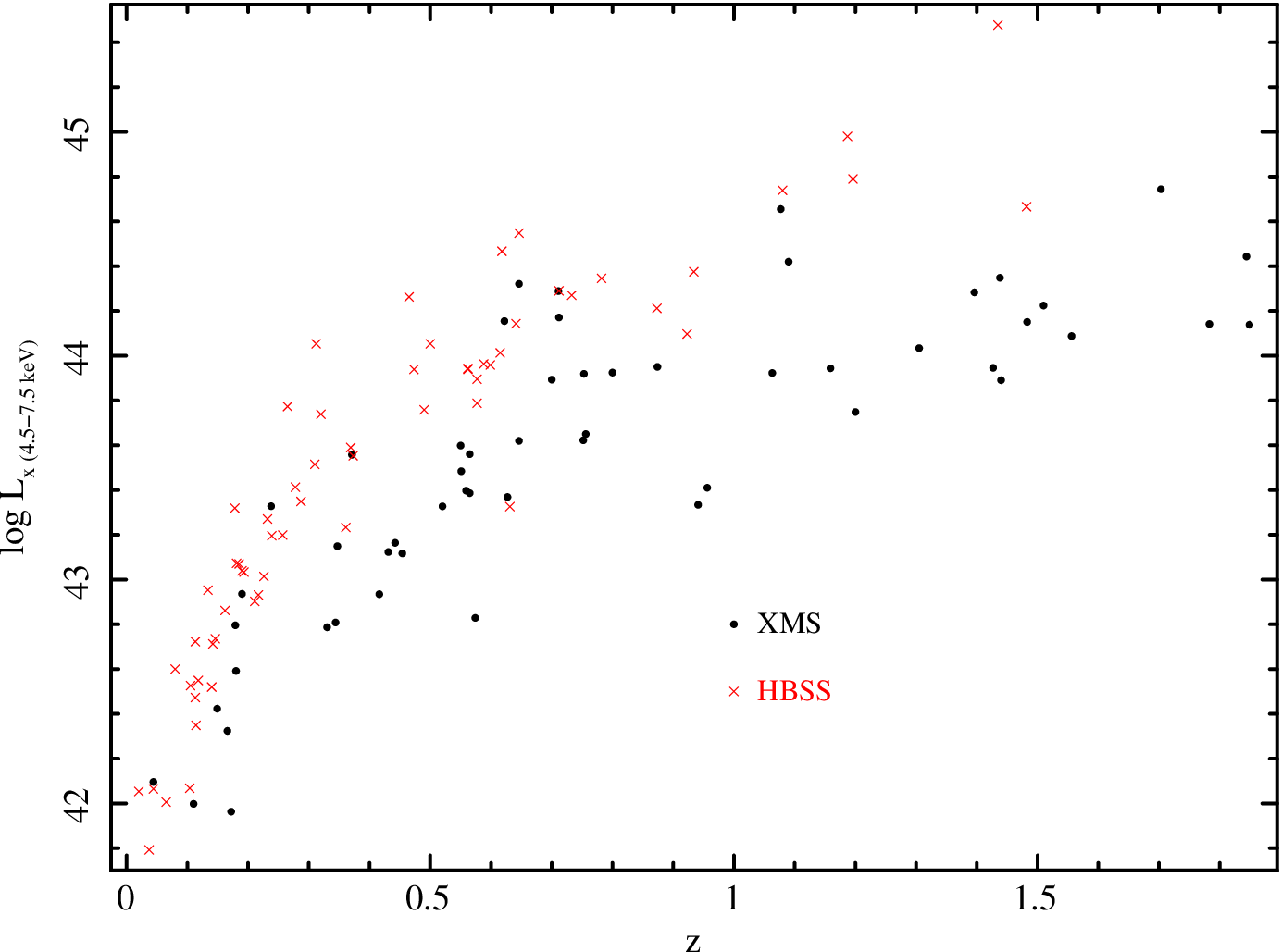}
}
\caption[]{Luminosity-redshift plane of different X-ray surveys in the Soft ({\it top panel}), Hard ({\it Center panel}) and Ultrahard ({\it Bottom panel}) bands.}
\label{fig:Lz}
\end{figure}

This paper is organized as follows: in section~\ref{Xdata} we summarize all the X-ray samples used in this work, giving a short overview and details on the level of completeness, sky areas, flux limits and energy bands in which they have been selected. In section~\ref{Nhfunction} we study the absorbing column density distribution of the spectroscopically identified AGN detected in hard X-rays ($>$2~keV) and we model the intrinsic fraction of absorbed AGN as a function of both luminosity and redshift in the 2-10 and 4.5-7.5~keV bands. In section~\ref{XLF} we compute the X-ray luminosity function of AGN in several energy bands using two methods: a modified version of the classic $1/V_a$ method to construct binned luminosity functions, and a Maximum Likelihood fit to an analytic model using all the sources available without binning. In section~\ref{discussion} we discuss the results obtained and we compare them with those obtained in previous works. Finally, the conclusions extracted from this work are reported in section~\ref{conclusions}. 

In what follows, X-ray luminosities and intrinsic column densities $N_H$ are in units of erg s$^{-1}$ and cm$^{-2}$, respectively. Throughout this paper we have assumed a cosmological framework with $H_0=70$~km s$^{-1}$ Mpc$^{-1}$, $\Omega_M=0.3$ and $\Omega_\Lambda=0.7$ (Spergel at al. \cite{Spergel03}). 

\section{The X-ray data}
\label{Xdata}

The backbone of the work presented in this paper is the XMS sample (Barcons et al. \cite{Barcons07}). This survey has typical exposure times of about 15 ks and hence is a survey reaching moderate depths, but in order to cover a wide luminosity and redshift range it is imperative to combine it with other complementary shallower and deeper surveys. Shallower, wider area surveys will provide significant numbers of bright sources at low redshifts, while deep pencil-beam surveys will probe fainter sources at greater distances.

Since the XMS survey is composed of different subsamples selected in several energy bands, we need to incorporate additional surveys for each of them. In the analysis performed in this paper we have used three energy bands defined as follows:

\begin{itemize}

\item {\it Soft}: 0.5-2~keV

\item {\it Hard}: 2-10~keV.

\item {\it Ultrahard}: 4.5-7.5~keV

\end{itemize}

\begin{figure}
\centering
\hbox{
\includegraphics[width=6.5cm,angle=270.0]{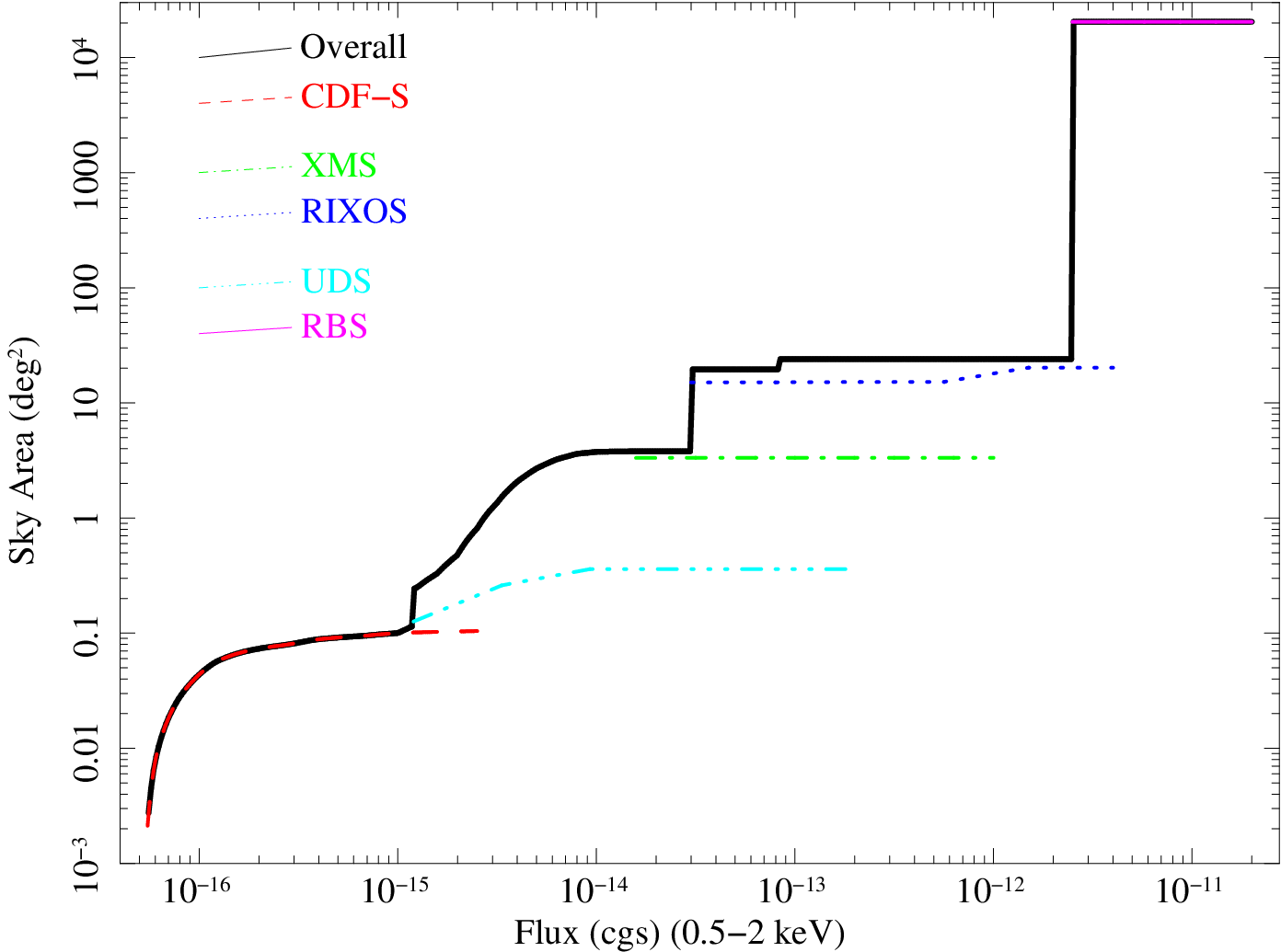}
}
\hbox{
\includegraphics[width=6.5cm,angle=270.0]{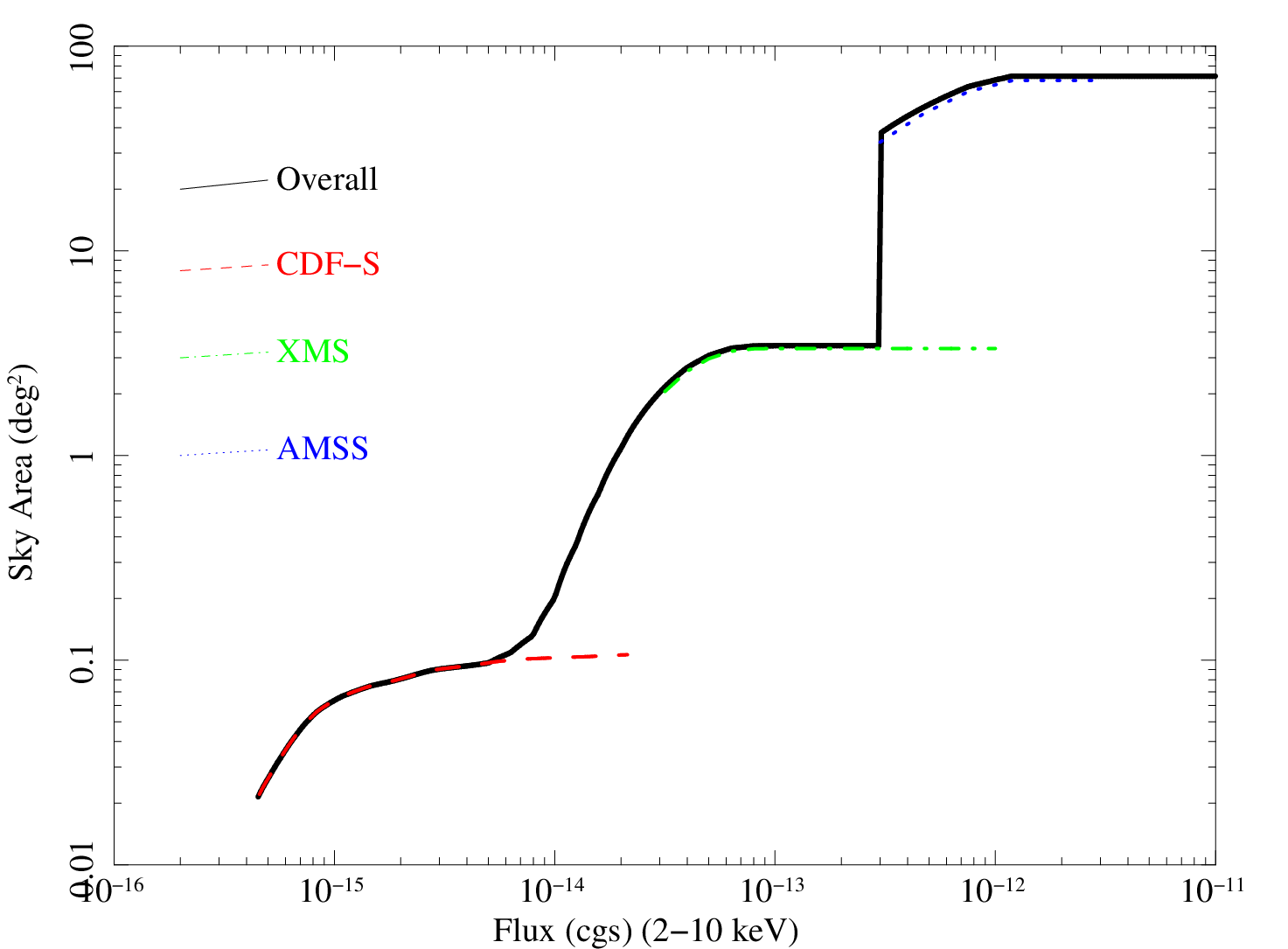}
}
\hbox{
\includegraphics[width=6.5cm,angle=270.0]{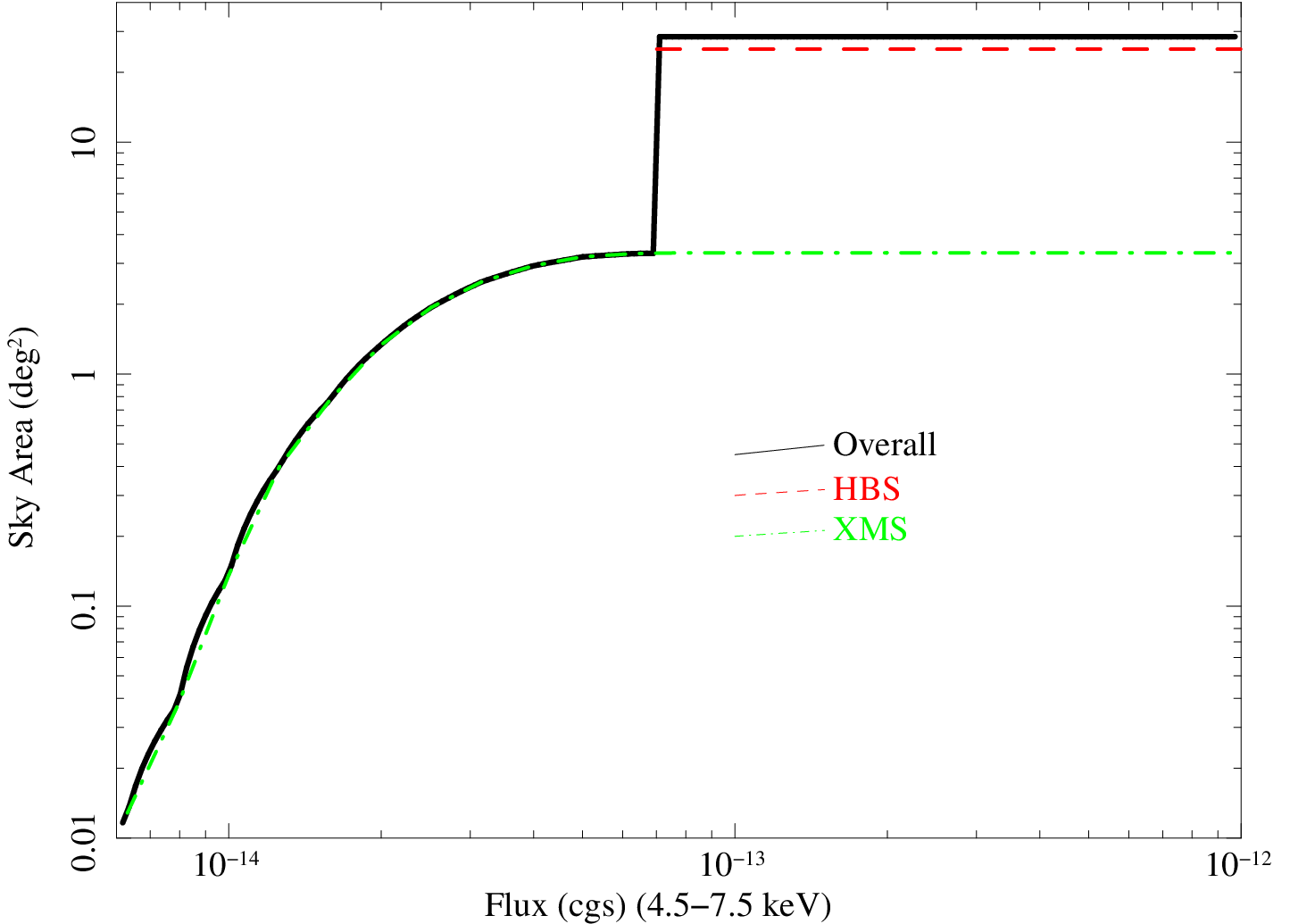}
}
\caption[]{Sky area of the survey as a function of flux in the Soft ({\it top panel}), Hard ({\it Center panel}) and Ultrahard ({\it Bottom panel}) bands.}
\label{fig:skyareas}
\end{figure}

In this section we will describe all the surveys used in this paper, which are also summarized in Table~\ref{tab:surveys}. They are all well-defined flux-limited surveys with very high identification completenesses, taken from already published catalogues of sources from past and present X-ray observatories (\XMM{}, \Chan{}, \ASCA{}, \ROSAT{}). In order to have an optimal $L_X-z$ plane coverage (see Figure~\ref{fig:Lz}), we will constrain our analysis to those sources identified as AGN with redshifts in the range $0.01 < z < 3$ in the Soft and Hard bands (spanning up to seven and six orders of magnitude in luminosity, respectively), and $0.01 < z < 2$ in the Ultrahard band (spanning up to four orders of magnitude in luminosity), being the vast majority of them spectroscopical redshifts. The total sky areas covered by these surveys are shown in Figure~\ref{fig:skyareas}. For the purposes of this work, and to avoid errors and biases caused by further classification, we have used the entire AGN population available (within the redshift limits stated above) irrespective of whether they were optically identified as Type-1 or Type-2 AGN.

\begin{table*}
  \centering
  \caption[]{Summary of surveys used in this work, along with their flux limits, sky coverage, total number of sources (identification completeness in parenthesis), and the number of identified AGN for each energy band.}
  \label{tab:surveys}

  \begin{tabular}{lcccc}
    \hline
    \hline
    \noalign{\smallskip}
        &   \multicolumn{2}{r}{Soft (0.5-2~keV)}   \\
    \noalign{\smallskip}
    \hline
    \noalign{\smallskip}
    Survey &  Flux limit (\cgs{}) & Area (deg$^{2}$) & $N_{total}$ & $N_{AGN}$ \\
    \noalign{\smallskip}
    \hline
    \noalign{\smallskip}
    RBS    &  2.5$\times$10$^{-12}$  & 20300 & 953 (100\%) & 310    \\
    RIXOS8 &  8.4$\times$10$^{-14}$  & 4.44 & 105 (100\%) &  40    \\
    RIXOS3 &  3.0$\times$10$^{-14}$  & 15.77 & 296 (94\%) & 182    \\
    XMS    &  1.5$\times$10$^{-14}$  & 3.33 & 210 (96\%) &  178    \\
    UDS    &  1.2$\times$10$^{-15}$  & 0.36 & 94 (95\%)$^a$ &  73$^a$    \\
    CDF-S  &  5.5$\times$10$^{-17}$  & 0.12 & 307 (99\%)$^a$ & 226$^a$    \\
    \noalign{\smallskip}
    \hline
    \hline
    \noalign{\smallskip}
        &   \multicolumn{2}{r}{Hard (2-10~keV)}  \\
    \noalign{\smallskip}
    \hline
    \noalign{\smallskip}
    Survey &  Flux limit (\cgs{}) & Area (deg$^{2}$) & $N_{total}$ & $N_{AGN}$ \\
    \noalign{\smallskip}
    \hline
    \noalign{\smallskip}
    AMSS   &  3.0$\times$10$^{-13}$  & 68 & 87 (99\%) &  79    \\
    XMS    &  3.3$\times$10$^{-14}$  & 3.33 & 159 (84\%) & 120    \\
    CDF-S  &  4.5$\times$10$^{-16}$  & 0.11 & 251 (99\%)$^a$ & 236$^a$    \\
    \noalign{\smallskip}
    \hline
    \hline
    \noalign{\smallskip}
        &    \multicolumn{2}{r}{Ultrahard (4.5-7.5~keV)}  \\
    \noalign{\smallskip}
    \hline
    \noalign{\smallskip}
    Survey &  Flux limit (\cgs{}) & Area (deg$^{2}$) & $N_{total}$ & $N_{AGN}$ \\
    \noalign{\smallskip}
    \hline
    \noalign{\smallskip}
    HBSS    &  7.0$\times$10$^{-14}$  & 25.17 & 67 (97\%) &  62    \\
    XMS    &  6.8$\times$10$^{-15}$  & 3.33 & 70 (86\%) &  57    \\
    \noalign{\smallskip}
    \hline
    \noalign{\smallskip}
    \multicolumn{2}{l}{$^a$ Including photometric redshifts (see text).}\\
  \end{tabular}
\end{table*}

\subsection{\XMM{} Medium Survey}
\label{XMS}

The \XMM{} Medium Sensitivity Survey (XMS, Barcons et al. \cite{Barcons07}) is a flux-limited survey of serendipitous X-ray sources at intermediate fluxes built from 25 \XMM{} fields of the AXIS sample (Carrera et al. \cite{Carrera07}). It covers a geometric sky area of 3.33 deg$^2$ and comprises 318 distinct X-ray sources, out of which 272 (86\%) have been spectroscopically identified using a number of ground-based facilities. This fraction, however, increases significatively in the Soft band, where the identification completeness is 96\%. The flux limit in the Soft band is 1.5$\times$10$^{-14}$~\cgs{}, while in the Hard and Ultrahard bands are 3.3$\times$10$^{-14}$ and 6.8$\times$10$^{-15}$~\cgs{}, respectively. In the Soft and Hard bands these limits are well above the sensitivity of the data, while in the Ultrahard band it corresponds to the flux of the faintest detected source in this sample (see Barcons et al. \cite{Barcons07} for details). For this work we have excluded the sources classified as stars, normal galaxies and clusters of galaxies, ending up with a sample of 178 AGNs in the Soft band, 120 in the Hard band and 57 in the Ultrahard band. A fraction of the objects classified as normal galaxies (which are $\sim$4\% over the total sample in the Soft and Ultrahard bands, and $\sim$5\% in the Hard band) might be optically elusive AGN given their high intrinsic luminosities ($\log L_X \gtrsim 42$). In these cases we have adopted a conservative position and excluded them from the final sample, both in the XMS and the other surveys involved.

\subsection{\XMM{} Hard Bright Sample}
\label{HBSS}

The \XMM{} Hard Bright Survey (HBSS) is part of a bigger survey project known as \XMM{} Bright Survey (XBS, Della Ceca et al. \cite{DellaCeca04}, Caccianiga et al. \cite{Caccianiga08}). In particular, the HBSS is a survey of sources at high Galactic latitudes detected in the 4.5-7.5~keV (Ultrahard) band down to a flux limit of 7$\times$10$^{-14}$~\cgs{}, with a flat sky coverage of 25.17 deg$^2$. The sample contains 67 sources. All of these sources but two have been spectroscopically identified, as reported in Caccianiga et al. (\cite{Caccianiga08}), with 62 of them being AGN. Along with the spectroscopic redshifts, information on the intrinsic absorption column densities for each source is also provided in Della Ceca et al. (\cite{DellaCeca08}) which will allow us to model the absorption of the sources detected in the Ultrahard band (most of them Type-2 AGN) as explained in Section~\ref{Nhfunction}.

\subsection{\Chan{} Deep Field South}
\label{CDFS}

The \Chan{} Deep Field South (CDF-S, Giacconi et al. \cite{Giacconi01}, Rosati et al. \cite{Rosati02}) is one of the deepest surveys in the Soft and Hard bands carried out so far, with a total exposure time of 1 Ms. The source samples are widely discussed in Bauer et al. (\cite{Bauer04}). We have used the 346 sources reported in Giacconi et al. (\cite{Giacconi02}) for a total of of 307 and 251 sources in the Soft and Hard bands, respectively, covering $\sim$0.125~deg$^2$ and $\sim$0.108 deg$^2$ down to flux limits of 5.5$\times$10$^{-17}$ and 4.5$\times$10$^{-16}$~\cgs{} in both energy bands. The optical imaging strategy and optical counterparts catalogue can be found in Giacconi et al. (\cite{Giacconi02}), while the spectroscopic identifications are taken from Szokoly et al. (\cite{Szokoly04}). In both cases the ground-based facilities used were the FORS1 and FORS2 instruments at the VLT. The redshifts of the inconclusive identifications and unidentified sources (which sum up to $\sim$60\% of the total CDF-S sample) have been taken from photometric redshifts estimations in Zheng et al (\cite{Zheng04}). These estimations make use of 12-band data in near ultraviolet, optical, infrared and X-rays along with sets of power law models for Type-1 AGN and various templates of galaxies. They have been calculated using two parallel models: HyperZ (Bolzonella, Miralles \& Pell\'o \cite{Bolzonella00}) and the Bayesian model BPZ (Ben\'itez \cite{Benitez00}). The photometric redshifts were checked against the secure spectroscopic identifications in Szokoly et al. (\cite{Szokoly04}) and the COMBO-17 survey (Wolf et al. \cite{Wolf01}, \cite{Wolf03}, \cite{Wolf04}) to ensure their reliability, matching well that of both surveys. The overall number of sources in Giacconi et al. (\cite{Giacconi02}) is 346, of which 145 have secure spectroscopic redshifts, 4 are unidentified and the remaining 197 sources have been photometrically identified. Photometric redshifts with flags 0.2, 0.3 and 0.4 in Zheng et al. (\cite{Zheng04}) are considered low-quality identifications since they rely only in one of the methods described in the paper (16 sources in total, 4.7\% of the total CDF-S sample). We have also made use of these 16 identifications in this work since their number is too low to significantly affect the XLF parameters given the total size of the samples used here. Since we need a highly complete survey at faint fluxes in order not to degrade the overall sample completeness and given the accuracy of the estimations, this is the only case in which we are using photometric redshifts in this work. The total number of sources classified as AGN in the CDF-S that we are using in this paper is 226 in the Soft band and 236 in the hard band. The $N_H$ column densities have been taken from Tozzi et al. (\cite{Tozzi06}).

\begin{figure*}
\centering
\hbox{
\includegraphics[width=6.5cm,angle=270.0]{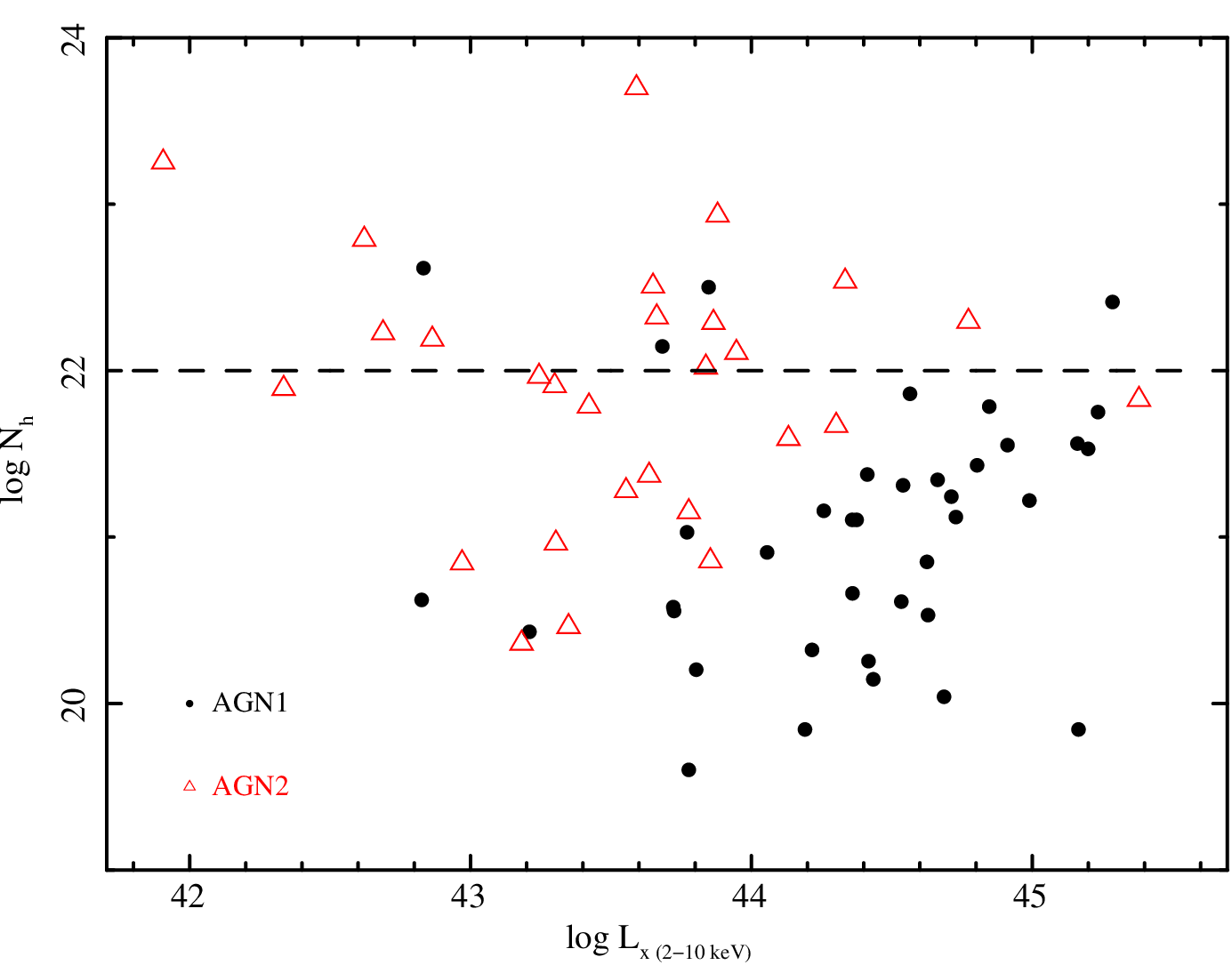}
\includegraphics[width=6.5cm,angle=270.0]{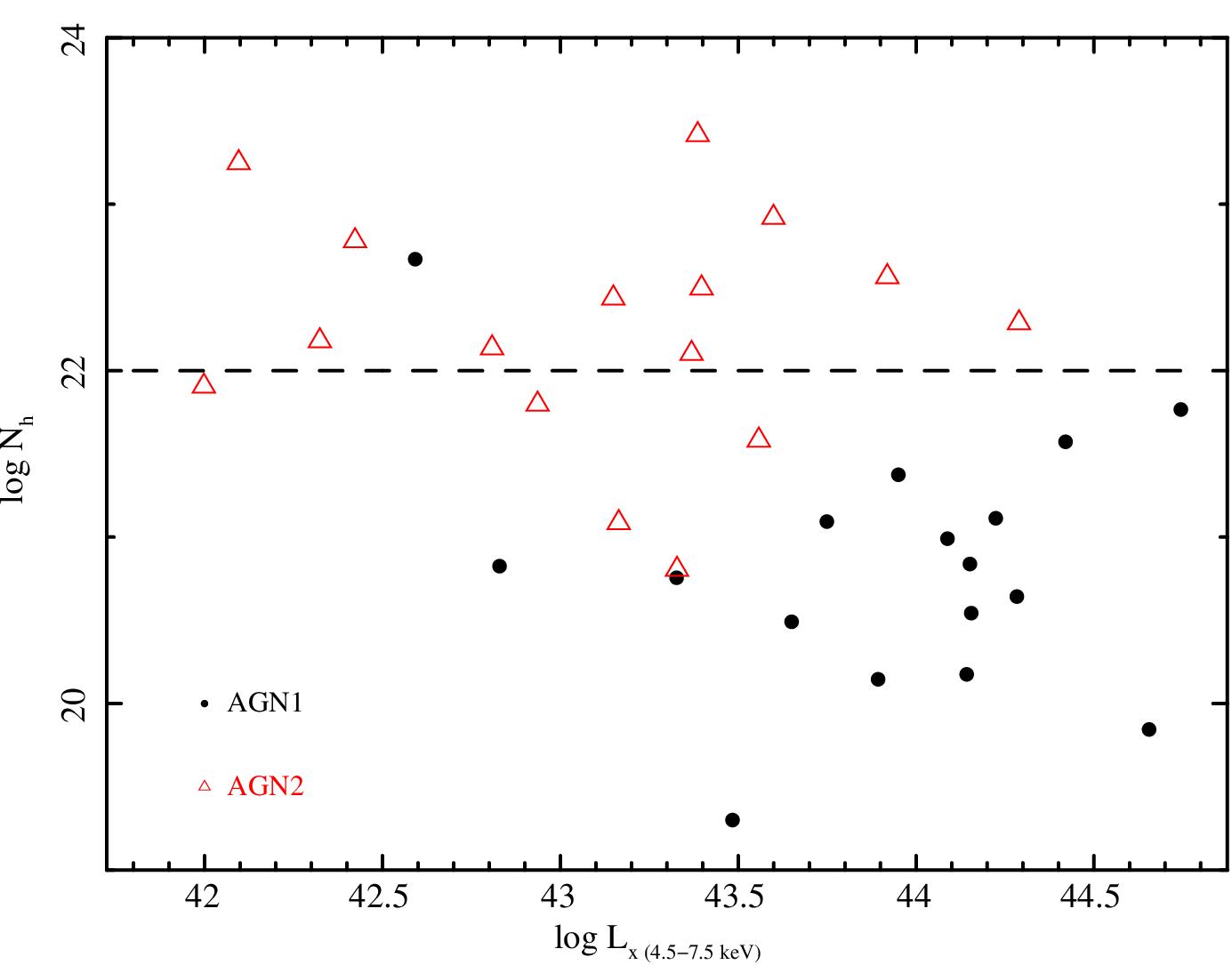}
}
\caption[]{Intrinsic $N_H$ {\it versus} X-ray luminosity of XMS sources detected in the Hard ({\it Left panel}) and Ultrahard ({\it Right panel}) bands. Dots represent those sources identified as Type-1 AGN while the Triangles are those classified as Type-2 AGN. The dashed line at $\log N_H = 22$ marks the standard separation between absorbed and unabsorbed AGN.}
\label{fig:NhLXMS}
\end{figure*}

\subsection{\ASCA{} Medium Sensitivity Survey}
\label{AMSS}

The \ASCA{} Medium Sensitivity Survey (AMSS, Ueda et al. \cite{Ueda05}) is one of the largest high Galactic latitude, broad band, X-ray surveys to date. It includes 606 sources in the 2-10~keV (Hard) band over a sky area of 278 deg$^2$ with hard band fluxes spanning between 10$^{-13}$ and 10$^{-11}$~\cgs{}. For this work we have used a flux-limited subsample of the AMSS in the northern sky (AMSSn). It has a total sky coverage of 68 deg$^2$, and all but one of the 87 hard X-ray sources detected down to a flux limit of 3$\times$10$^{-13}$~\cgs{} have been spectroscopically identified (79 of them being AGN). The $N_H$ measurements have been taken from Table~2 in Akiyama et al. (\cite{Akiyama03}). Further details on the imaging and spectroscopic identifications can also be found in Akiyama et al. (\cite{Akiyama03}).

\subsection{\ROSAT{} International X-ray/Optical Survey}
\label{RIXOS}

The \ROSAT{} International X-ray/Optical Survey (RIXOS, Mason et al. \cite{Mason00}) is a medium sensitivity survey of high Galactic latitude X-ray sources. The sample contains 401 sources divided in two subsamples: 64 \ROSAT{} fields with 296 sources down to a flux limit of 3$\times$10$^{-14}$~\cgs{} (known as RIXOS3), plus 18 further fields containing 105 sources down to a flux limit of 8.4$\times$10$^{-14}$~\cgs{} (RIXOS8) in the Soft band. RIXOS3 covers 15.77 deg$^2$ in the sky while RIXOS8 covers 4.44 deg$^2$, summing a total sky area of $\sim$20.2 deg$^2$. The RIXOS3 subsample has been spectroscopically identified up to a rate of 94\% (with total population of 182 AGN), whereas the RIXOS8 subsample has been completely identified (adding 40 AGN to the full sample), which makes them ideal for evolution studies in the Soft band.

\subsection{\ROSAT{} Deep Survey - Lockman Hole}
\label{RDS}

The \ROSAT{} Deep Survey (RDS, Hasinger et al. \cite{Hasinger98}) collects all the observations performed by \ROSAT{} in the period 1990-1997 in the direction of the Lockman Hole, which is one of the areas of the sky with a minimum of the Galactic Hydrogen column density (Lockman et al. \cite{Lockman86}). In this paper we use an extension of the RDS, the \ROSAT{} Ultra Deep Survey (UDS, Lehmann et al. \cite{Lehmann01}), which comprises 94 X-ray sources down to a flux limit of 1.2$\times$10$^{-15}$~\cgs{} ($\sim$5 times fainter than the RDS) in the Soft band. This sample is 90\% spectroscopically identified, containing 70 AGN (mainly Type-1) plus 3 photometrically identified Type-2 AGN that have been also included in the analysis for a total of 73 AGN.

%

\begin{figure*}
\centering
\hbox{
\includegraphics[width=6.5cm,angle=270.0]{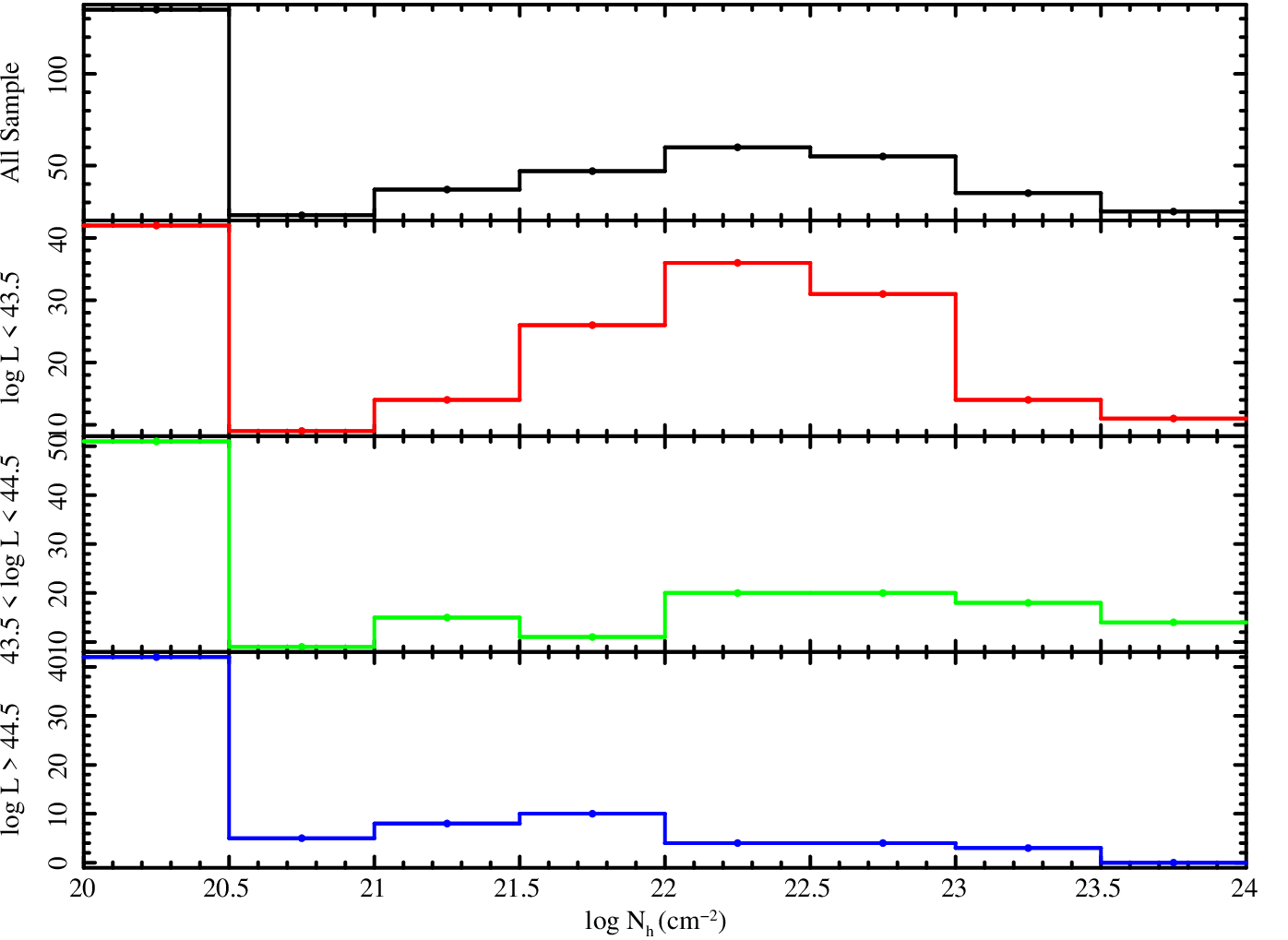}
\includegraphics[width=6.5cm,angle=270.0]{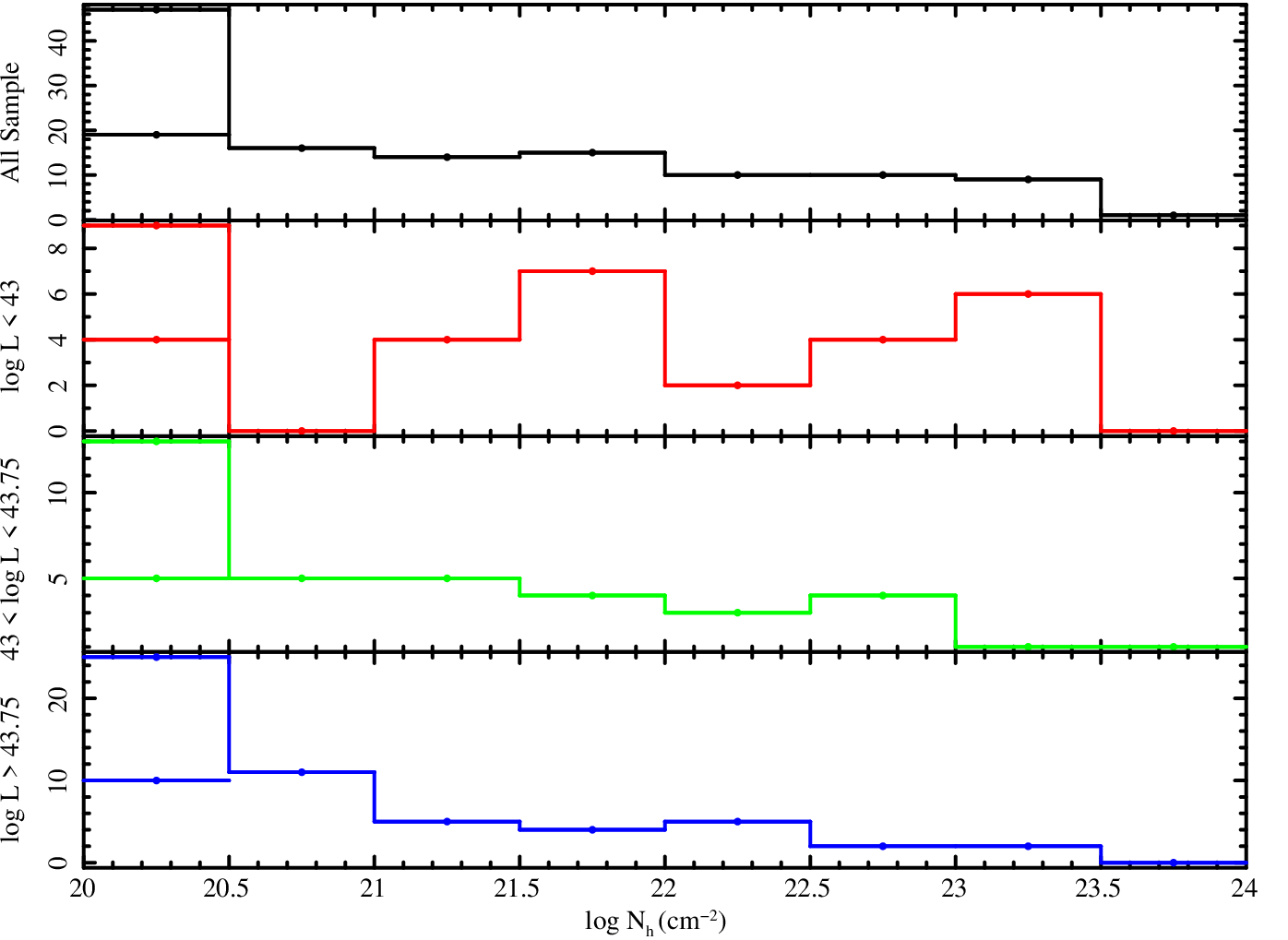}
}
\caption[]{Observed $N_H$ distributions of the joint XMS/AMSS/CDF-S sample in the Hard band ({\it Left panel}) and XMS/HBSS sample in the Ultrahard band ({\it Right panel}), respectively. The distribution is shown for the full sample and in different luminosity ranges. The first bin accounts for both the sources observed in that bin (lower step) and these sources plus the sources with $\log N_H < 20$ or uncomputed absorptions (higher step).}
\label{fig:Nhhist}
\end{figure*}


\subsection{\ROSAT{} Bright Survey}
\label{RBS}

The \ROSAT{} Bright Survey (RBS, Fischer et al. \cite{Fischer98}, Schwope et al. \cite{Schwope00}) aimed to identify the brightest $\sim$2000 sources detected in the \ROSAT{} All Sky Survey (RASS, Voges et al. \cite{Voges99}) at high Galactic latitudes, excluding the Magellanic Clouds and the Virgo cluster, with PSPC count rates above 0.2 s$^{-1}$. This sample contains bright sources (flux limit $\sim$2.5$\times$10$^{-12}$~\cgs{} in the Soft band) with redshifts mainly below 1 and a total sky coverage of $\sim$20300 deg$^2$. The sample has been completely identified via spectroscopic observations and includes 310 AGN, plus a strong population of galaxy clusters which has been excluded from this work.


\section{The $N_H$ function of hard sources}
\label{Nhfunction}

This work aims at calculating the cosmological evolution of the X-ray luminosity function of all AGN (both unabsorbed and absorbed) within our sample in three energy bands. This should be done by calculating the intrinsic (before absorption) luminosity and absorption ($N_H$) function in order to obtain results free from selection effects. Since hard X-ray photons are less affected by absorption, absorbed AGN are more likely to be detected at energies above 2~keV and hence we have applied such method to the Hard (2-10~keV) and Ultrahard (4.5-7.5~keV) sources, for which we have detailed spectral data (photon index $\Gamma$ and intrinsic absorbing column densities $N_H$, Mateos et al. \cite{Mateos05}). With $\sim$25\% of the XMS sources in the Hard and Ultrahard bands classified as Type-2 AGN, we find that they typically $\log N_H > 22$, a value commonly used to separate absorbed AGN from the unabsorbed ones (see Figure~\ref{fig:NhLXMS}). An appropiate calculation of the X-ray luminosity function of these sources would therefore require a previous modelling of the intrinsic absorption in order to avoid possible selection effects (see Della Ceca et al. \cite{DellaCeca08} for a discussion on this subject). On the contrary, the sources detected in the Soft band (0.5-2~keV) are so strongly affected by absorption that, in this band, we can assume that we are effectively sampling mainly unabsorbed AGN.

The $N_H$ function $f(L_X,z;N_H)$ is a probability distribution function for the absorption column density as a function of the X-ray luminosity and redshift. It is measured in units of $(\log N_H)^{-1}$ and is normalized to unity over a defined $N_H$ region:

\begin{equation}
\label{eq:Nhnorm}
\int_{\log N_{H_{min}}}^{\log N_{H_{max}}} f(L_X,z;N_H) d \log N_H = 1
\end{equation}

\begin{figure*}
\centering
\hbox{
\includegraphics[width=6.5cm,angle=270.0]{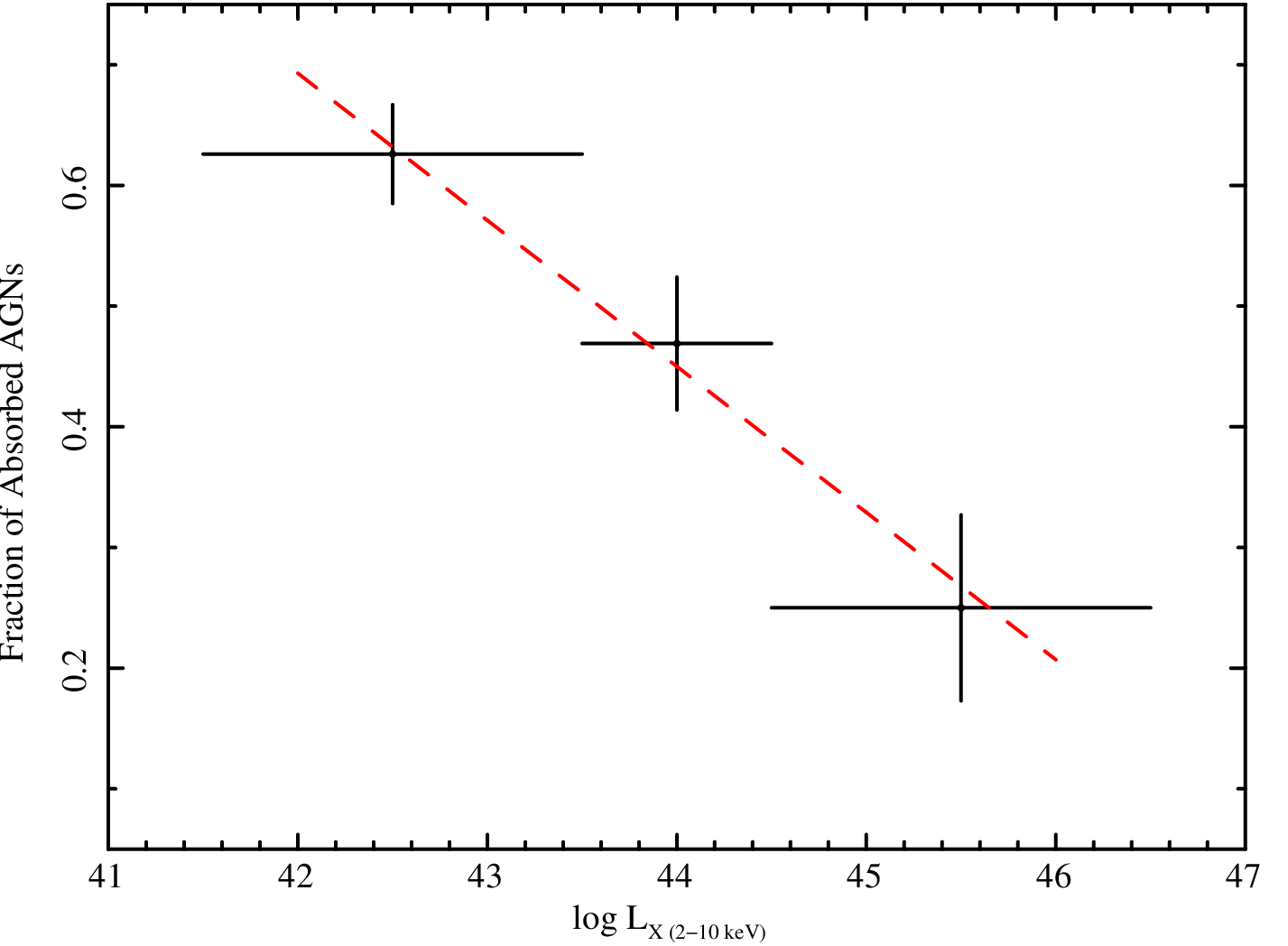}
\includegraphics[width=6.5cm,angle=270.0]{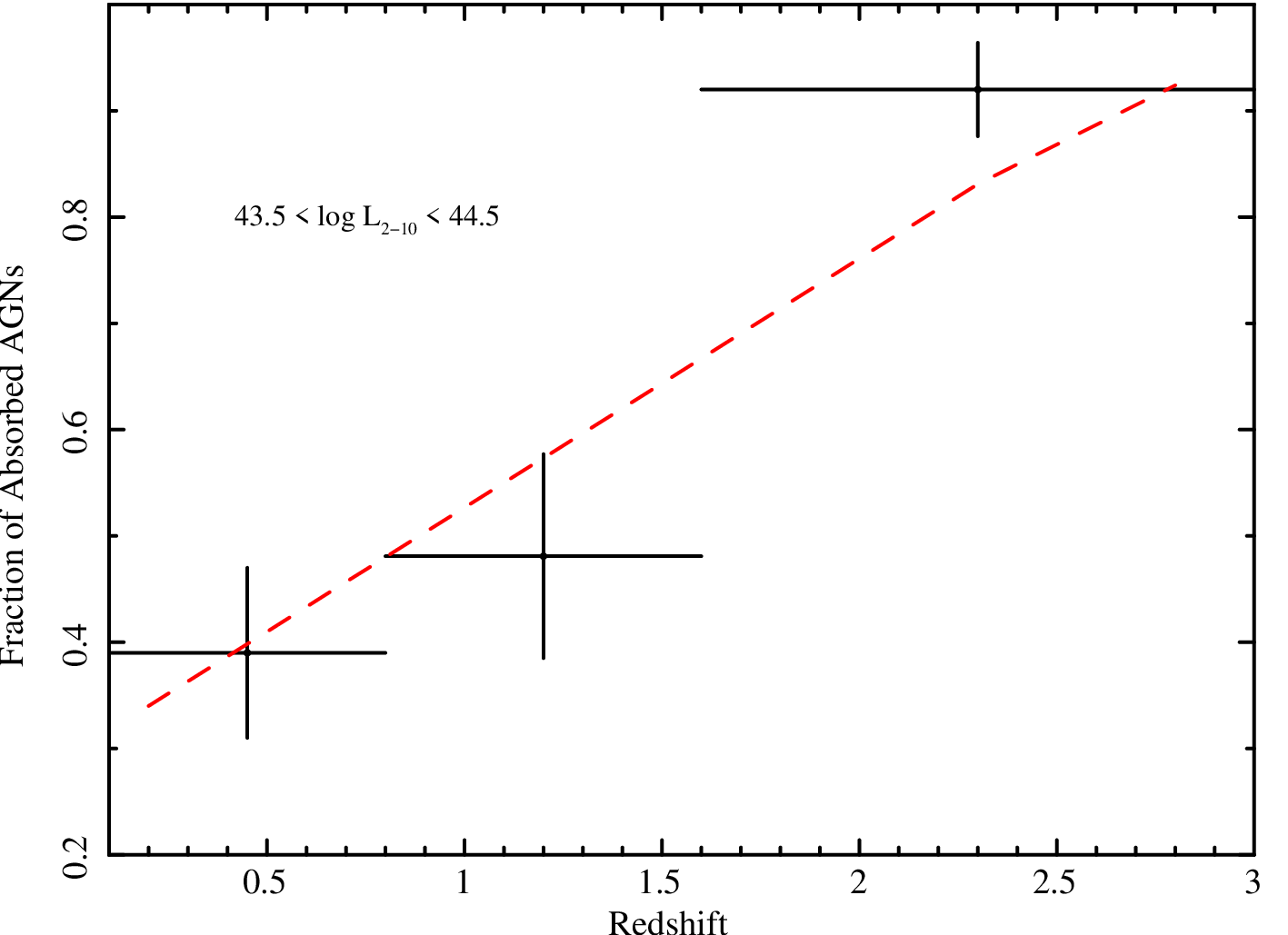}
}
\caption[]{{\it Left panel:} Fraction of absorbed AGN as a function of the 2-10~keV luminosity. {\it Right panel:} Fraction of absorbed AGN as a function of redshift in the luminosity range $\log L_X = 43.5-44.5$. Dashed lines represent the best-fit model of the $N_H$ function using $N_H=10^{22}$~cm$^{-2}$ as the dividing value between obscured and unobscured AGN.}
\label{fig:absfrac_hard}
\end{figure*}

\begin{figure*}
\centering
\hbox{
\includegraphics[width=6.5cm,angle=270.0]{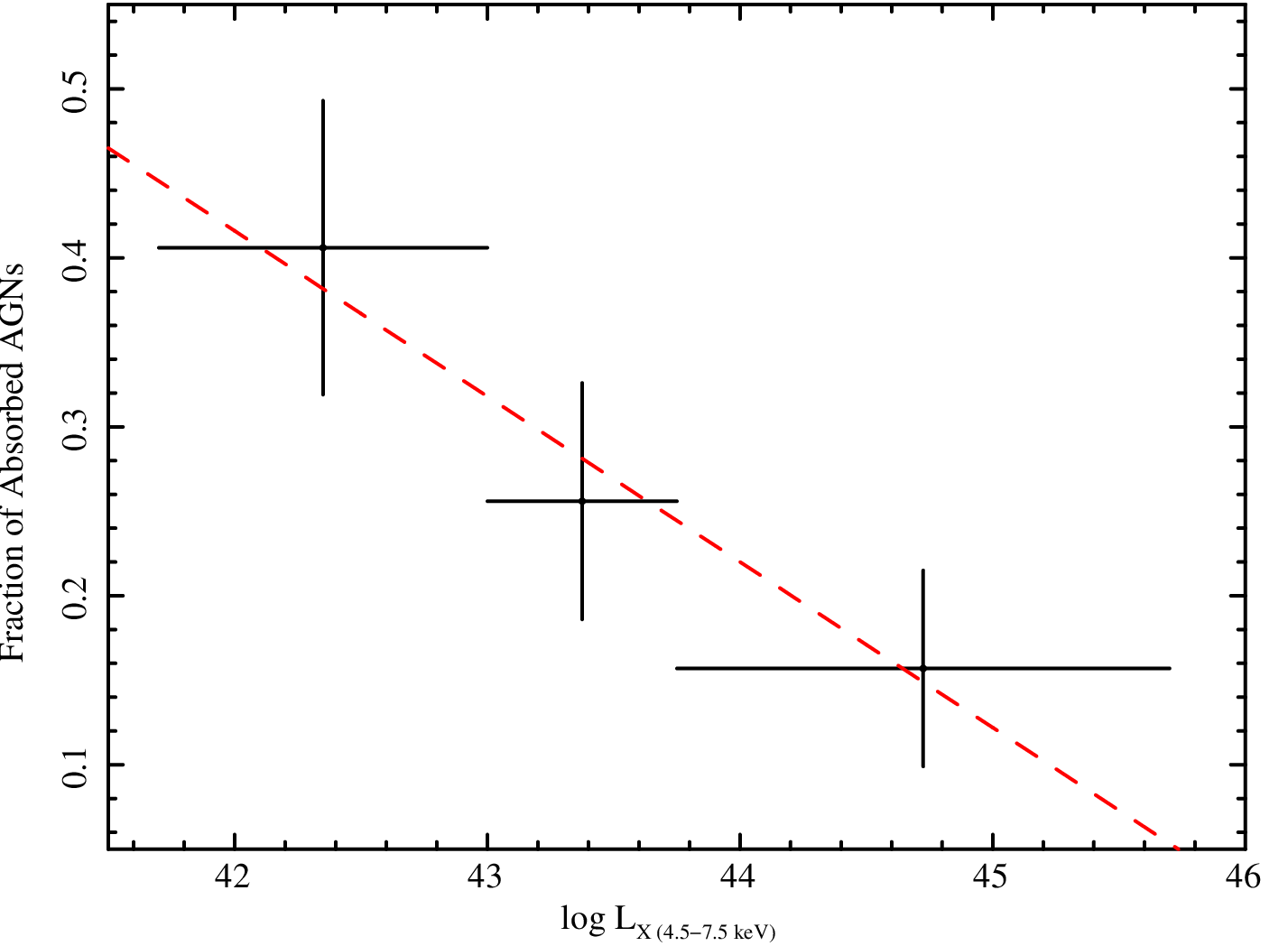}
\includegraphics[width=6.5cm,angle=270.0]{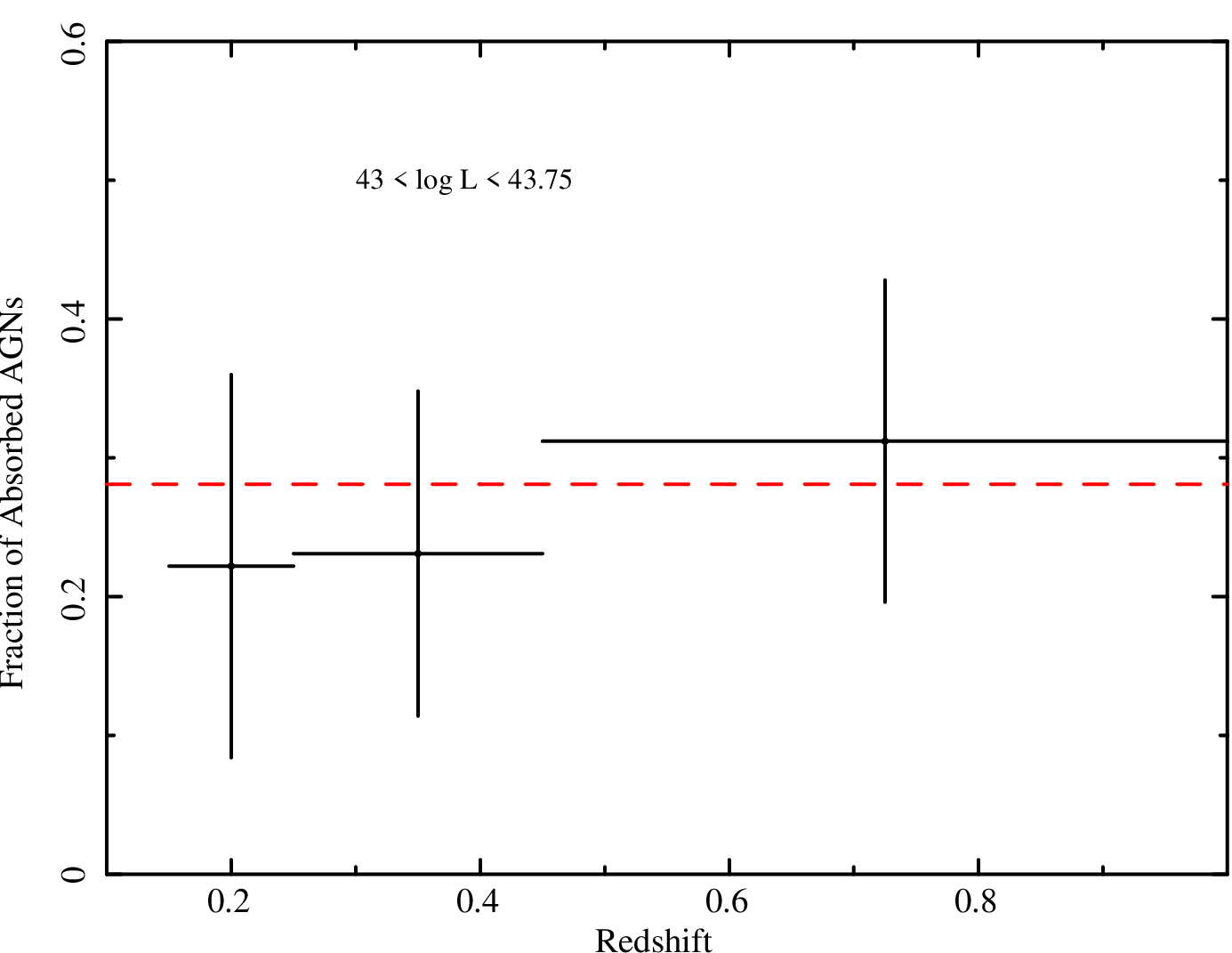}
}
\caption[]{{\it Left panel:} Fraction of absorbed AGN as a function of the 4.5-7.5~keV luminosity. {\it Right panel:} Fraction of absorbed AGN as a function of redshift in the luminosity range $\log L_X = 43-43.75$. Dashed lines represent the best-fit model of the $N_H$ function using $N_H=10^{22}$~cm$^{-2}$ as the dividing value between obscured and unobscured AGN.}
\label{fig:absfrac_uh}
\end{figure*}

\noindent We have chosen the $N_H$ limits to be $\log N_{H_{min}} = 20$ and $\log N_{H_{max}} = 24$, given the intrinsic $N_H$ range spanned by our joint samples (see Figure~\ref{fig:Nhhist}). The observed fraction of absorbed AGN (those with $\log N_H > 22$) is measured by the parameter $\psi$, which is in general function of both luminosity and redshift. If we compare the value of $\psi$ in different luminosity ranges we can observe that the fraction of absorbed AGN is not constant, decreasing as the luminosity increases in both the Hard and Ultrahard bands. The fraction of absorbed AGN clearly increases with redshift in the case of the Hard band (see Figure~\ref{fig:absfrac_hard}), mainly due to the sources above $z\gtrsim 2$ coming from the CDF-S survey. If we pay attention only to the points below $z\sim 2$, the evolution in redshift looks milder. On the other hand, there is not significant variation in the value of $\psi$ with redshift at a given luminosity in the Ultrahard band (see Figure~\ref{fig:absfrac_uh}), which can be partly explained by the poor coverage in redshift of this sample. Therefore, we will assume that the formal expression of $\psi$ is dependent on both the X-ray luminosity and redshift and hence we have formalized its parametrization as a linear function of $\log L_X$ and $z$, similarly as in La Franca et al. \cite{LaFranca05}:

\begin{equation}
\label{eq:psi}
\psi(L_X,z)=\psi_{44}\left[(\log L_X - 44)\beta_L + 1\right]\left[(z - 0.5)\beta_z + 1\right]
\end{equation}

\noindent where $\psi_{44}$ is the fraction of absorbed AGN at $\log L_X = 44$ and $z = 0.5$, and $\beta_L$ and $\beta_z$ are the slopes of the linear dependencies on luminosity and redshift, respectively.

Given equation~\ref{eq:Nhnorm}, the normalized $N_H$ function can be written as:

\begin{equation}
\label{eq:Nhfunc}
f(L_X,z;N_H)=\left\{
\begin{array}{lc}
\frac{1-\psi(L_X,z)}{2} & ; 20 \leq \log N_H < 22 \\
\frac{\psi(L_X,z)}{2} & ; 22 \leq \log N_H \leq 24 \\
\end{array}
\right\}
\end{equation}

In order to obtain the best-fit values of the free parameters $\psi_{44}$, $\beta_L$ and $\beta_z$ we have performed a $\chi^2$ fit on the sources of the Hard and Ultrahard samples, although in the latter case we have fixed $\beta_z = 0$ to account for the absence of dependence on redshift observed in this band. The values thus obtained are $\psi_{44}=0.41_{-0.04}^{+0.03}$, $\beta_L=-0.22_{-0.05}^{+0.04}$ and $\beta_z=0.57_{-0.10}^{+0.12}$ for the Hard band, and $\psi_{44}=0.22\pm0.04$ and $\beta_L=-0.45_{-0.25}^{+0.20}$ for the Ultrahard band (see Table~\ref{tab:XLF}). The 1$\sigma$ errors correspond to $\Delta\chi^2=1$.

We find that our best-fit value at $\log L_{2-10} = 44$ ($\psi_{44} = 0.41_{-0.04}^{+0.03}$) is in excellent agreement with the ones calculated by Ueda et al. (\cite{Ueda03}) ($\psi_{44}=0.41\pm0.03$) and La Franca et al. (\cite{LaFranca05}) ($\psi_{44} = 0.42_{-0.04}^{+0.03}$). Hasinger (\cite{Hasinger08}) found a strong linear decrease from $\sim$80\% to $\sim$20\% in the range $\log L_X=42-46$. A linear fit to the data in Table~5 of Hasinger (\cite{Hasinger08}) yields a best-fit value at $\log L_{2-10} = 44$ of $0.38\pm0.04$, and a slope of $-0.226\pm0.014$, also in agreement with our results in the 2-10~keV band within the error bars. The decrease in the absorbed AGN fraction at high luminosities have been reported by many authors (Ueda et al. \cite{Ueda03}, La Franca et al. \cite{LaFranca05}, Akylas et al. \cite{Akylas06}, Della Ceca et al. \cite{DellaCeca08}).

La Franca et al. (\cite{LaFranca05}) reported that the absorbed fraction is also dependent on the redshift. Redshift dependence is also found by Treister \& Urry (\cite{Treister06}) and Ballantyne et al. (\cite{Ballantyne06}), while Dwelly \& Page (\cite{Dwelly06}) did not find dependence on either the luminosity or redshift. In a recent paper, Hasinger (\cite{Hasinger08}) has found a strong decrease in the fraction of absorbed AGN with X-ray luminosity and a significant increase of that fraction with redshift. The result of Hasinger (\cite{Hasinger08}) suggests that the evolution of this fraction at a fixed luminosity of $\log L_{2-10} = 43.75$ is faster than the result obtained by Treister \& Urry (\cite{Treister06}) (probably due to the fact that the latter used the Broad Line AGN classification only, which tends to overestimate the fraction of absorbed AGN at low redshifts) and is consistent with our best fit model at a typical redshift of $z=0.5$ ($\psi(z=0.5)=0.37\pm0.07$ against our prediction of $\psi(z=0.5)=0.39\pm0.08$).

The lack of any dependence on redshift of our Ultrahard sample, however, must be handled with caution since it comes from two similar \XMM{} surveys only and spans a limited redshift range. As stated in Perola et al. (\cite{Perola04}), who found no luminosity dependence from a sample of 117 sources in the 2-10~keV band, deeper X-ray surveys are needed to take into account those sources at higher redshifts and lower luminosities in order to fully investigate the true incidence of absorption. Only combining several surveys at different depths, as we have done in the Hard band, and therefore covering wider regions in the $L_X-z$ plane is possible to unveil the real dependencies.

In a recent work by Della Ceca et al. (\cite{DellaCeca08}) the dependency on luminosity of the fraction of absorbed AGN detected in the 4.5-7.5~keV band is calculated using $N_H=4\times10^{21}$~cm$^{-2}$ as the dividing value between obscured and unobscured AGN. This corresponds to $A_V\sim 2$~mag for a given Galactic $A_V/N_H$ ratio of $5.27\times10^{22}$~mag cm$^{-2}$ which is the best dividing value between optical Type-1 AGN and optical Type-2 AGN in the HBSS sample as shown in Caccianiga et al. (\cite{Caccianiga08}). If we repeat the calculations in the Ultrahard band using this $N_H$ value as a dividing line we obtain $\psi_{44}=0.30_{-0.05}^{+0.04}$ and $\beta_L=-0.42_{-0.18}^{+0.15}$ as the best fit parameters for the $N_H$ function. They are slightly different but consistent within the 1$\sigma$ error bars to those calculated with a dividing value of $N_H=10^{22}$~cm$^{-2}$.

Della Ceca et al. (\cite{DellaCeca08}) found a fraction of absorbed AGN with $L_{2-10}\geq 3 \times 10^{42}$~erg s$^{-1}$ of $0.57\pm0.11$ which was in excellent agreement with the results obtained by a variety of {\it SWIFT} and {\it INTEGRAL} surveys (see Table 3 in Della Ceca et al. \cite{DellaCeca08}). The value predicted by our Ultrahard $N_H$ function, converting 2-10~keV luminosities to 4.5-7.5~keV using $\Gamma=1.7$, is $0.55\pm0.18$ using a dividing value of $N_H=4\times10^{21}$~cm$^{-2}$, which fully agrees with that of Della Ceca et al. (\cite{DellaCeca08}). Considering a dividing value of $N_H=10^{22}$~cm$^{-2}$, our predicted fraction decreases to $0.42\pm0.18$ which coincides with the {\it INTEGRAL} result of Sazonov et al. (\cite{Sazonov07}) ($0.42\pm0.09$) who used the same dividing $N_H$ value between absorbed and unabsorbed AGN.



\section{The X-ray Luminosity Function}
\label{XLF}

In this section we will calculate the X-ray luminosity function (XLF) of our sources in the Soft, Hard and Ultrahard bands. The differential XLF measures the number of AGN per unit of comoving volume $V$ and $\log L_X$, and is a function of both luminosity and redshift:

\begin{equation}
\label{eq:diffXLF}
\frac{d\Phi(L_X,z)}{d\log L_X} = \frac{d^2N(L_X,z)}{dV d\log L_X}
\end{equation}

\noindent and it is assumed that is a continous function over the luminosity and redshift ranges where it is defined.

As a first approach we will estimate the XLF in fixed luminosity and redshift bins (section~\ref{binnedXLF}) which will allow us to have a general overview of the overall XLF behavior. In section~\ref{modelXLF} we will express the XLF in terms of an analytical formula over which we will perform a Maximum Likelihood (ML) fit, using the full information available from each single source and thus avoiding biases coming from finite bin widths and from the imperfect sampling of the $L_X-z$ plane.

\subsection{Binned luminosity function}
\label{binnedXLF}

There are a variety of methods to estimate the binned XLF. The classical approach is the $1/V_a$ method (Schmidt \cite{Schmidt68}), which was later generalized by Avni \& Bahcall (\cite{Avni80}) for samples with multiple flux limits. This method has been widely used to compute the binned XLF in evolution studies of flux-limited samples (i.e. Maccacaro et al. \cite{Maccacaro91}, Della Ceca et al. \cite{DellaCeca92}, Ellis et al. \cite{Ellis96}) but it can lead to systematic errors, especially when dealing with sources very close to the flux limit. A number of techniques have been proposed to solve this problem (Page \& Carrera \cite{Page00}, Miyaji et al. \cite{Miyaji01}). In this work, we have used the alternative method proposed by Page \& Carrera (\cite{Page00}) to calculate the differential binned luminosity function, which avoids most of the biases that accompany the classic $1/V_a$ method.

The differential XLF can be approximated by:

\begin{equation}
\label{eq:binnedXLF}
\frac{d\Phi(L_X,z)}{d\log L_X}\approx\frac{N}{\int_{L_{min}}^{L_{max}}\int_{z_{min}}^{z_{max}(L)}\frac{dV}{dz}dzd\log L_X}
\end{equation}

\noindent where $N$ is the number of objects found in a given luminosity-redshift bin $\Delta L \Delta z$ which is surveyed by the double integral in the denominator. The differential comoving volume has been calculated using the expression from Hogg (\cite{Hogg99}):

\begin{equation}
\label{eq:Va}
\frac{dV}{dz} = \frac{c}{H_0}\left(\frac{D_L}{1+z}\right)^2\left(\Omega_M(1+z)^3+\Omega_{\Lambda}\right)^{-1/2}\Omega(L_X,z)
\end{equation}

\noindent where $D_L$ is the luminosity distance and $\Omega(L_X,z)$ is the solid angle subtended by the survey. Note that some portion of the $\Delta L \Delta z$ bin may represent objects that are fainter than the survey flux limit. The upper limit of the integral in redshift $z_{max}(L)$ is hence either the top of the redshift shell $\Delta z$ or the redshift at the survey flux limit for a given luminosity: $z_{max}(L)=z_{lim}(L_X,S_{lim})$.

If the sample is composed of multiple flux-limited surveys (as it happens in this work), they are added 'coherently' as explained in Avni \& Bahcall (\cite{Avni80}). This means we will assume that each object could be found in any of the survey areas for which is brighter than the corresponding flux limit.

The binned XLF thus obtained for our sample presents an overall double power-law shape, with a steeper slope at brighter luminosities beyond a given break luminosity that takes place somewhere in the range $\log L_X = 43-44$. The position of this break seems to change with redshift, moving to higher values of $\log L_X$ as we go deeper, thus showing clear evidences of evolution in the AGN that compose our sample. The shape at higher redshifts is less constrained due to the limited statistics. An analytical model is therefore needed to account for these changes in shape as it evolves with $z$, and whose parameters could be compared with previous works.

\subsection{Analytical model}
\label{modelXLF}

In the light of the results obtained in section~\ref{binnedXLF}, we will express the XLF as an analytic function with an smooth behavior over the range of $L_X$ and $z$ under study in this work. The parameters of this function will describe its overall shape and how it evolves with luminosity and redshift, which will express physical properties of the population under study.

As shown in many works in the Soft and Hard bands (Miyaji et al. \cite{Miyaji00}, Ueda et al. \cite{Ueda03}, Hasinger et al. \cite{Hasinger05}, Silverman et al. \cite{Silverman08}), the XLF seems to be best described as a double power law modified by a factor for evolution. Here we implement two different models: the simpler Pure Luminosity Evolution (PLE) model in which the X-ray luminosity evolves with redshift, and the more complicated Luminosity-dependent Density Evolution (LDDE) model in which the evolution factor not only depends on the redshift but also on the luminosity. Both models can be expressed respectively as:

\begin{equation}
\label{eq:PLEmodel}
\frac{d\Phi(L_X,z)}{d\log L_X}=\frac{d\Phi(L_X/e(z),0)}{d\log L_X}
\end{equation}

\noindent for the PLE model, and:

\begin{equation}
\label{eq:LDDEmodel}
\frac{d\Phi(L_X,z)}{d\log L_X}=\frac{d\Phi(L_X,0)}{d\log L_X}e(z,L_X)
\end{equation}

\noindent for the LDDE model. The first factor of the second term in equations~\ref{eq:PLEmodel} and~\ref{eq:LDDEmodel} describes the shape of the present-day XLF for which we adopt a smoothly connected double power law:

\begin{equation}
\label{dphi0}
\frac{d\Phi(L_X,0)}{d\log L_X}=A\left[\left(\frac{L_X}{L_0}\right)^{\gamma_1}+\left(\frac{L_X}{L_0}\right)^{\gamma_2}\right]^{-1}
\end{equation}

\noindent where $\gamma_1$ and $\gamma_2$ are the slopes, $L_0$ is the value of the luminosity where the change of slope occurs, and $A$ is the normalization.

The evolution factor of the PLE model is expressed as:

\begin{equation}
\label{eq:ez}
e(z)=\left\{
\begin{array}{lc}
(1+z)^{p_1}  &  ; z<z_c \\
e(z_c)\left(\frac{1+z}{1+z_c}\right)^{p_2} & ; z\geq z_c \\
\end{array}
\right\}
\end{equation}

\noindent where the parameters $p_1$ and $p_2$ account for the evolution below and above, respectively, the cut-off redshift $z_c$. In the LDDE model we will assume that the cut-off redshift depends on the X-ray luminosity:

\begin{equation}
\label{eq:ezL}
e(z,L_X)=\left\{
\begin{array}{lc}
(1+z)^{p_1}  &  ; z<z_c(L_X) \\
e(z_c)\left(\frac{1+z}{1+z_c(L_X)}\right)^{p_2} & ; z\geq z_c(L_X) \\
\end{array}
\right\}
\end{equation}

\begin{equation}
\label{eq:zc}
z_c(L_X)=\left\{
\begin{array}{lc}
z_c^* & ; L_X\geq L_a \\
z_c^*\left(\frac{L_X}{L_a}\right)^{\alpha} & ; L_X<L_a \\
\end{array}
\right\}
\end{equation}

where $\alpha$ measures the strength of the dependence of $z_c$ with luminosity.

\begin{table*}
  \centering
  \caption[]{Parameters of the X-ray luminosity function.}
  \label{tab:XLF}

  \begin{tabular}{lcccccc}
    \hline
    \hline
    \noalign{\smallskip}
        & \multicolumn{2}{c}{Soft (0.5-2~keV)}  & \multicolumn{2}{c}{Hard (2-10~keV)} & \multicolumn{2}{c}{Ultrahard (4.5-7.5~keV)} \\
    \noalign{\smallskip}
    \hline
    \hline
    \noalign{\smallskip}
      &  PLE & LDDE &  PLE & LDDE &  PLE & LDDE \\
    \noalign{\smallskip}
    \hline
    \noalign{\smallskip}
        &   &  \multicolumn{3}{c}{Present day XLF parameters} \\
    \noalign{\smallskip}
    \hline
    \noalign{\smallskip}
    $A^a$          & 50.73$\pm$3.79 &  3.76$\pm$0.38 & 17.96$_{-6.09}^{+9.97}$ &  4.78$_{-0.23}^{+0.20}$ & 10.41$_{-1.46}^{+1.87}$ & 1.32$\pm$0.20    \\
    $\log L_0$$^b$  & 42.90$_{-0.02}^{+0.03}$ &  43.56$\pm$0.05 & 43.60$\pm$0.13 &  43.91$_{-0.02}^{+0.01}$ & 43.00$_{-0.21}^{+0.18}$ & 43.43$_{-0.32}^{+0.23}$    \\
    $\gamma_1$     & 0.34$\pm$0.01 &  0.72$\pm$0.02  & 0.81$\pm$0.06 &  0.96$\pm$0.02 & 1.14$\pm$0.01 & 1.28$_{-0.16}^{+0.08}$    \\
    $\gamma_2$     & 2.01$\pm$0.04 &  2.04$\pm$0.04  & 2.37$_{-0.18}^{+0.19}$ &  2.35$\pm$0.07 & 2.72$_{-0.28}^{+0.29}$ &  2.53$\pm$0.32 \\
    \noalign{\smallskip}
    \hline
    \hline
    \noalign{\smallskip}
        &  &  \multicolumn{3}{c}{Evolution parameters}  \\
    \noalign{\smallskip}
    \hline
    \noalign{\smallskip}
    $p_1$          & 1.78$\pm$0.06 & 3.38$\pm$0.09 & 2.04$_{-0.12}^{+0.13}$ & 4.07$_{-0.07}^{+0.06}$ & 3.03$_{-0.22}^{+0.17}$ &  6.46$_{-0.29}^{+0.69}$    \\
    $p_2$          & 0.00 $(fixed)$ & -1.5 $(fixed)$ & 0.00 $(fixed)$ &  -1.5 $(fixed)$ & 0.00 $(fixed)$ & -1.5 $(fixed)$  \\
    $z_c$          & 1.7 $(fixed)$ & 1.42 $(fixed)$ & 1.9 $(fixed)$ & 1.9 $(fixed)$  & 1.9 $(fixed)$ & 1.9 $(fixed)$   \\
    $\log L_a$$^b$  & ... & 44.6 $(fixed)$ & ... & 44.6 $(fixed)$ & ... & 44.6 $(fixed)$  \\
    $\alpha$       & ... & 0.100$\pm$0.005 & ... &  0.245$\pm$0.003  & ... &  0.245 $(fixed)$   \\
    \noalign{\smallskip}
    \hline
    \hline
    \noalign{\smallskip}
        &  &  \multicolumn{3}{c}{$N_H$ function parameters}  \\
    \noalign{\smallskip}
    \hline
    \noalign{\smallskip}
    $\psi_{44}$ &  \multicolumn{2}{c}{...}  &  \multicolumn{2}{c}{0.41$_{-0.04}^{+0.03}$} & \multicolumn{2}{c}{0.22$\pm$0.04}    \\
    $\beta_{L}$ &  \multicolumn{2}{c}{...}  &  \multicolumn{2}{c}{-0.22$_{-0.05}^{+0.04}$} & \multicolumn{2}{c}{-0.45$_{-0.25}^{+0.20}$}   \\
    $\beta_{z}$ &  \multicolumn{2}{c}{...}  &  \multicolumn{2}{c}{0.57$_{-0.10}^{+0.12}$}  & \multicolumn{2}{c}{0.00 $(fixed)$}  \\
    \noalign{\smallskip}
    \hline
    \noalign{\smallskip}
    $P_{2DKS}(L_X,z)$$^c$ &  2$\times$10$^{-4}$  & 0.21   & 3$\times$10$^{-4}$  & 0.18   & 0.28   & 0.84  \\
    \hline
    \noalign{\smallskip}
    \multicolumn{2}{l}{$^a$ In units of 10$^{-6}h_{70}^3$~Mpc$^{-3}$.}\\
    \multicolumn{2}{l}{$^b$ In units of $h_{70}^{-2}$~erg s$^{-1}$.}\\
    \multicolumn{2}{l}{$^c$ 2D K-S test probability.}\\
  \end{tabular}
\end{table*}

\subsubsection{Model fitting to Soft sources}
\label{XLFsofthard}

We have fitted our sample in the Soft band to the PLE and LDDE models described above using a Maximum Likelihood (ML) method, which optimally exploits the information from each source without binning and is therefore free from all the biases commented in section~\ref{binnedXLF}. In this band we are sampling mainly unabsorbed AGN and therefore the derived XLF is that of the unabsorbed population in the 0.5-2~keV band.

The ML technique we have implemented here is that of Marshall et al. (\cite{Marshall83}), in which the likelihood function is defined as the product of the probabilities of observing exactly one object in the differential element $dzd\log L_X$ at each $(z_i,L_{X_i})$ for the $N$ objects in the sample, and of the probabilities of observing zero objects in all other differential elements in the accessible regions of the $z-L_X$ plane. The expression to be minimized is hence:

\begin{equation}
\label{eq:MLsh}
\begin{array}{lr}
S=-2\sum_{i=1}^Nln\frac{d\Phi(L_X^i,z^i)}{d\log L_X} &  \\
  +2\sum_{j=1}^{N_{sur}}\int_{z_1}^{z_2}\int_{L_1}^{L_2}\frac{d\Phi(L_X,z)}{d\log L_X}C^j(L_X,z)\frac{dV^j(L_X,z)}{dz}dzd\log L_X\\
\end{array}
\end{equation}

\noindent Here, the index $i$ runs over all the sources present in the sample whereas the index $j$ runs over all the surveys that compose the sample. In case there is incompleteness in the spectroscopic identifications in the $j$-th survey, the factor $C^j(L_X,z)$ accounts for the completeness of the identifications at a given X-ray flux. Here we assume that the redshift distribution of the unidentified sources is the same as that of the identified sources at similar fluxes and hence the effective survey area is the geometric area multiplied by this factor. Note that this assumption is not correct when the source is unidentified by non-random effects, but we do not expect this to affect our results significatively given the high degree of completeness achieved by our combined sample. The integrals are calculated over the full redshift ($0.01 < z < 3$) and luminosity ($40 \lesssim \log L_X < 46$) ranges spanned by our sample in the Soft band taking into account the flux limits of the different surveys that compose the sample. There are 5 sources with luminosities in the range ($40 \lesssim \log L_X < 42$) that have been classified as AGN (see Szokoly et al. \cite{Szokoly04}). Although they could be normal galaxies misclassified as AGN, we have decided to trust the original classification and include them in the final sample. The remaining sources labeled as AGN in this range come from the photometric classification in Zheng et al. (\cite{Zheng04}).

The expression for $S$ is minimized using the MINUIT software package (James \cite{James94}) from the CERN Program Library. 1$\sigma$ errors for each parameter are calculated by fixing the parameter of interest at different values and leaving the other parameters to float freely until $\Delta S$=1. Since the model is rather complex  and some of the parameters are unconstrained when performing the fit using the 9 free parameters, we have fixed some of them. In particular we have fixed $p_2$, $z_c$ and $L_a$ to the values obtained by Hasinger et al. (\cite{Hasinger05}) in the 0.5-2~keV band. This left us with 6 free parameters to fix: $A$, $\gamma_1$, $\gamma_2$, $L_0$, $p_1$ and $\alpha$ (see Table~\ref{tab:XLF}).

\begin{figure*}
\centering
\hbox{
\includegraphics[width=6.5cm,angle=270.0]{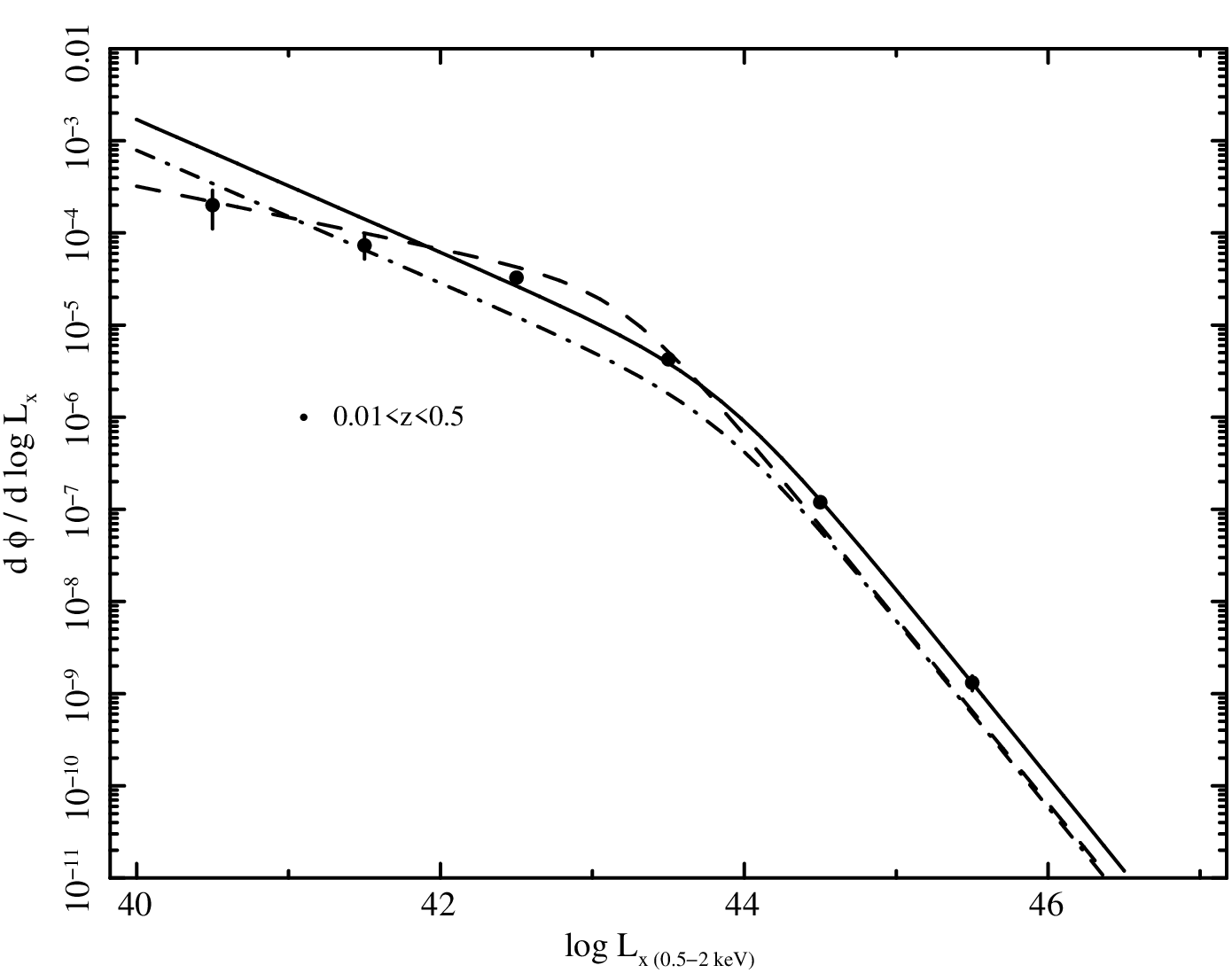}
\includegraphics[width=6.5cm,angle=270.0]{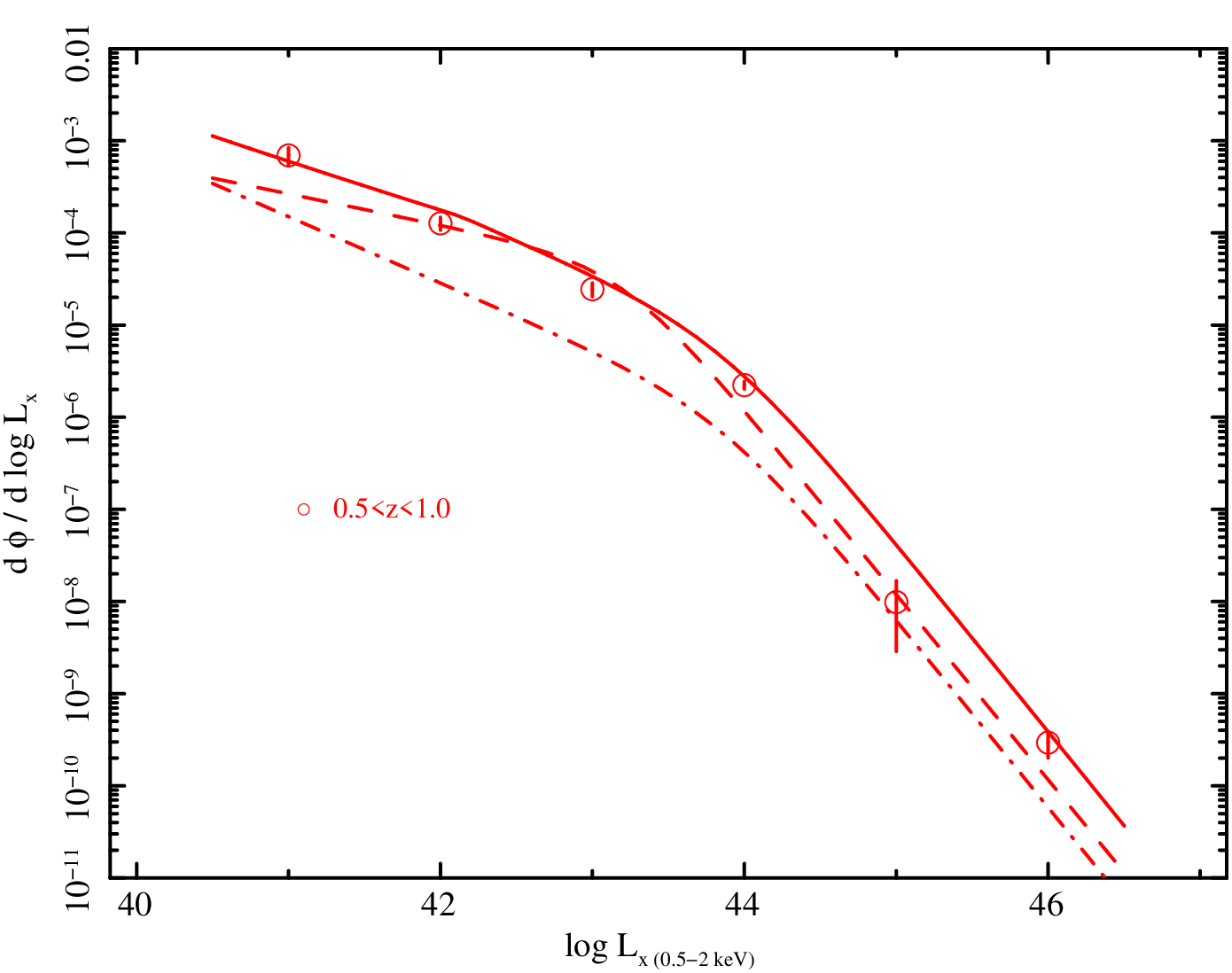}
}
\hbox{
\includegraphics[width=6.5cm,angle=270.0]{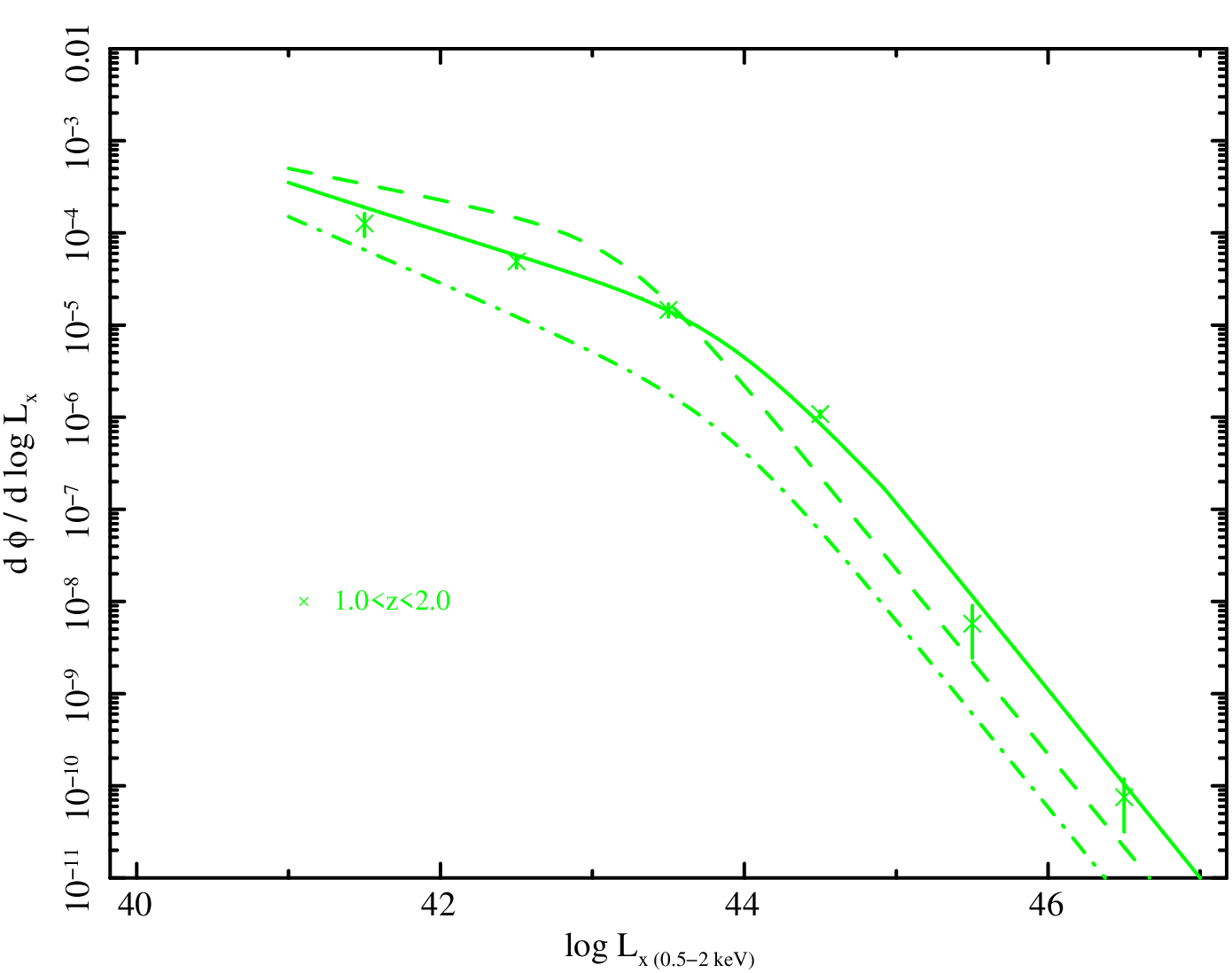}
\includegraphics[width=6.5cm,angle=270.0]{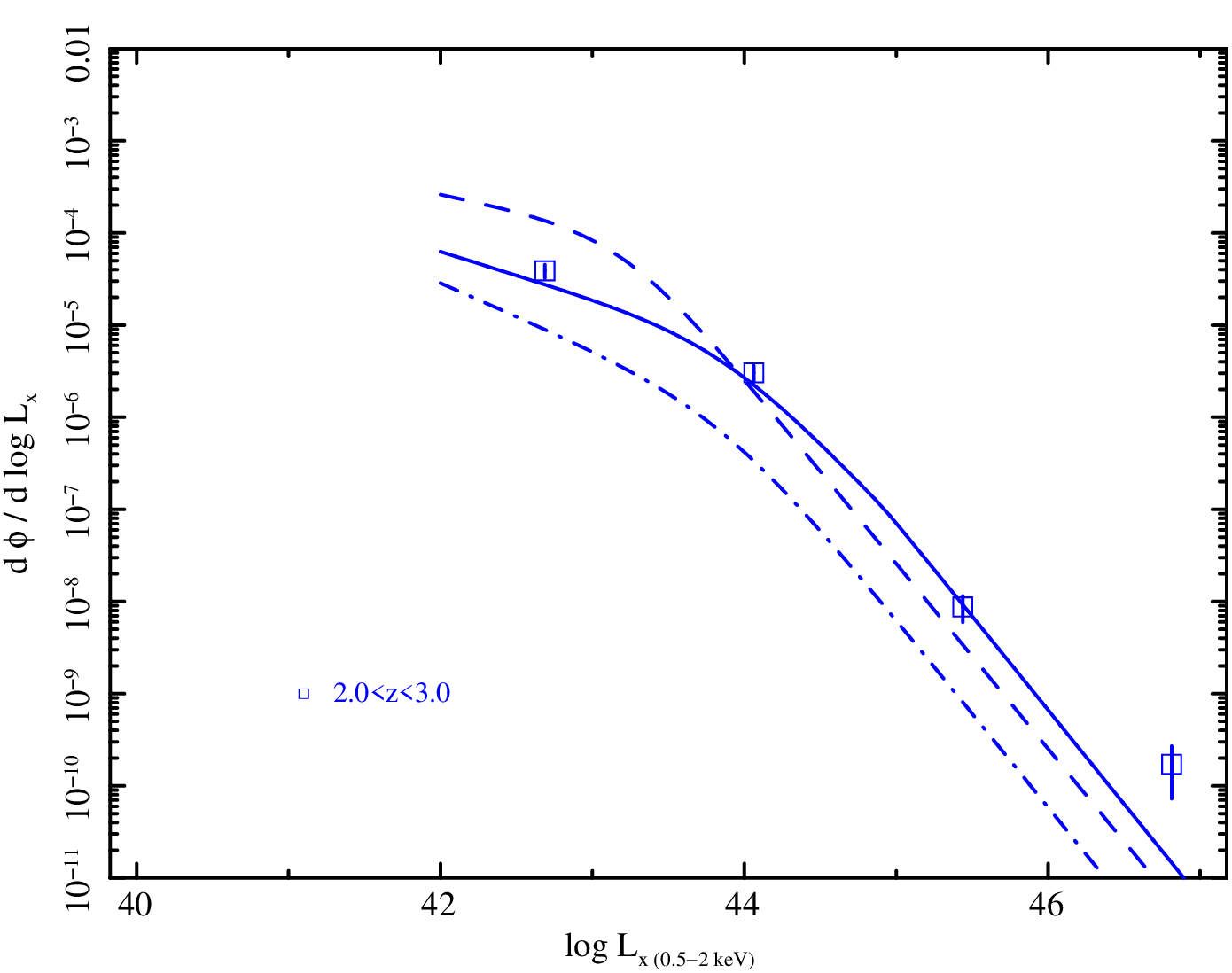}
}
\caption[]{X-ray luminosity function in the Soft band in the redshift ranges 0.01-0.5 ({\it Top left panel}), 0.5-1.0 ({\it Top right panel}), 1.0-2.0 ({\it Bottom left panel}) and 2.0-3.0 ({\it Bottom right panel}). The points are the results from the binned XLF and the lines are the best-fit curves to our analytical LDDE ({\it solid line}) and PLE ({\it dashed line}) models evaluated at the median of each redshift range. The dot-dashed line is our LDDE model evaluated at $z=0$.}
\label{fig:XLFsoft}
\end{figure*}

\subsubsection{Model fitting to Hard and Ultrahard sources}
\label{XLFuh}

For the sources in the Hard and Ultrahard bands we will make use of the information obtained in section~\ref{Nhfunction} when performing the ML fit. Given the coverage of the three-dimensional space $L_X-z-N_H$ spanned by our Hard and Ultrahard samples, we were not able to fit simultaneously the $N_H$ function and the XLF (which would lead us to a functional form with 11 free parameters). Instead, we will add the $N_H$ function to the analytic expression to be minimized but fixing their parameters, $\psi_{44}$, $\beta_L$ and $\beta_z$, to those obtained in section~\ref{Nhfunction}. Taking this into account, the expression for $S$ is:

\begin{equation}
\label{eq:MLuh}
\begin{array}{lr}
S=-2\sum_{i=1}^Nln\frac{d\Phi(L_X^i,z^i)}{d\log L_X}f(L_X^i,z^i;N_H^i) &  \\
  +2\sum_{j=1}^{N_{sur}}\int_{z_1}^{z_2}\int_{L_1}^{L_2}\int_{N_{H_1}}^{N_{H_2}}f(L_X,z;N_H)\frac{d\Phi(L_X,z)}{d\log L_X} \\
\times C^j(L_X,z,N_H)\frac{dV^j(L_X,z,N_H)}{dz}dzd\log L_Xd\log N_H \\
\end{array}
\end{equation}

Again, the integrals are calculated over the full $L_X-z-N_H$ space in the ranges $0.01 < z < 2$, $20 < \log N_H < 24$, $40 \lesssim \log L_{X(2-10)}\lesssim 46$ (in the Hard band) and $42 \lesssim \log L_{X(4.5-7.5)}< 45$ (in the Ultrahard band). The factor $C^j(L_X,z,N_H)$ accounts for the incompleteness of the identifications and $N_H$ measurements. Similarly as in the Soft band, there are 2 sources with $\log L_{X(2-10)} < 41$ that were originally classified as AGN which we have decided to keep in the final sample. As we did in the previous section, we fixed the XLF parameters $p_2$, $z_c$ and $L_a$ to those obtained by Ueda et al. (\cite{Ueda03}) for sources detected in the Hard band and, in addition, given the poor coverage of the $L_x-z$ plane achieved by the Ultrahard sample, we have also fixed the strength of the dependence of $z_c$ on luminosity $\alpha$ to the value we obtained when fitting the Hard sample.


\section{Discussion of the results}
\label{discussion}

\subsection{Comparison with other works}
\label{otherworks}

\begin{figure*}
\centering
\hbox{
\includegraphics[width=6.5cm,angle=270.0]{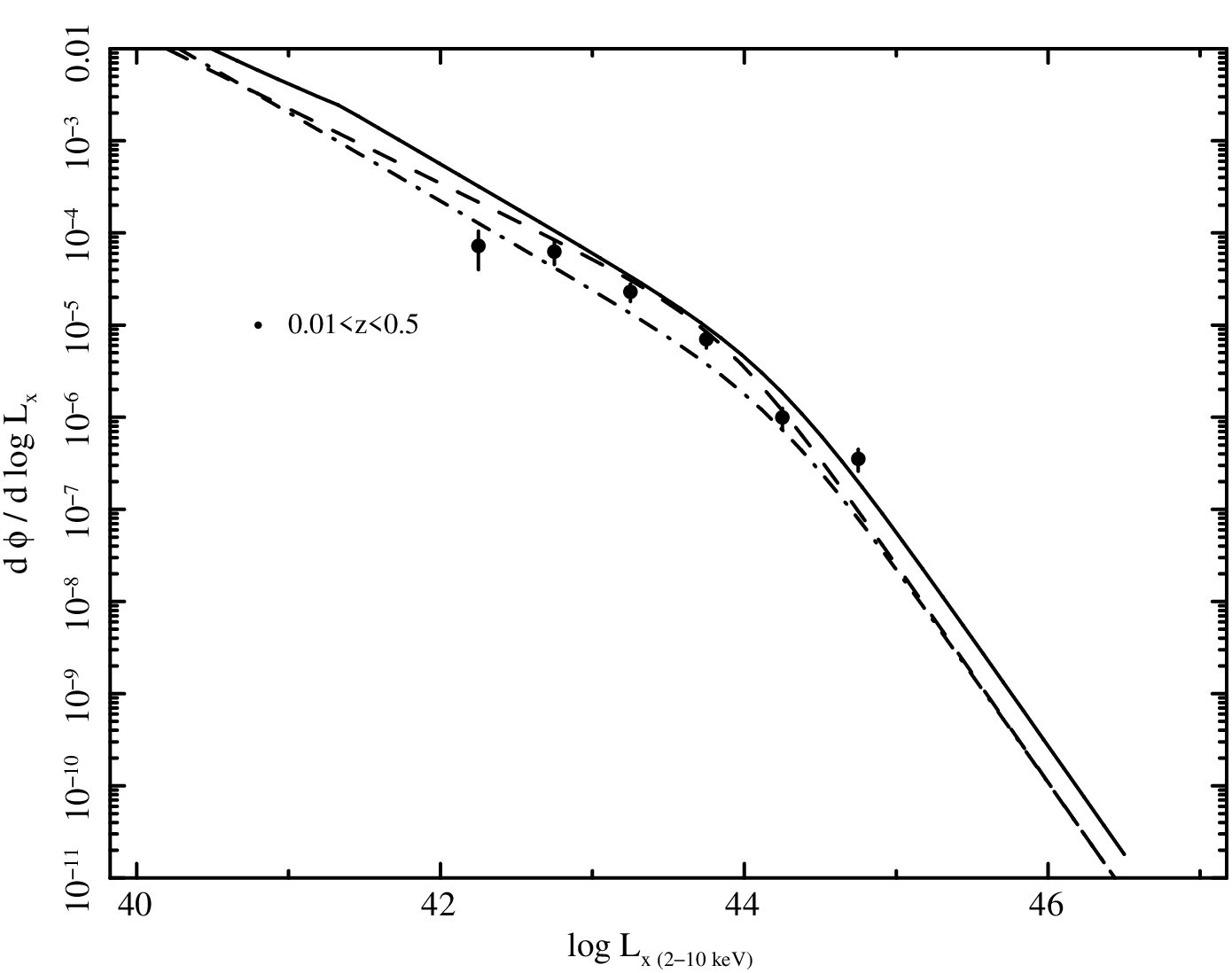}
\includegraphics[width=6.5cm,angle=270.0]{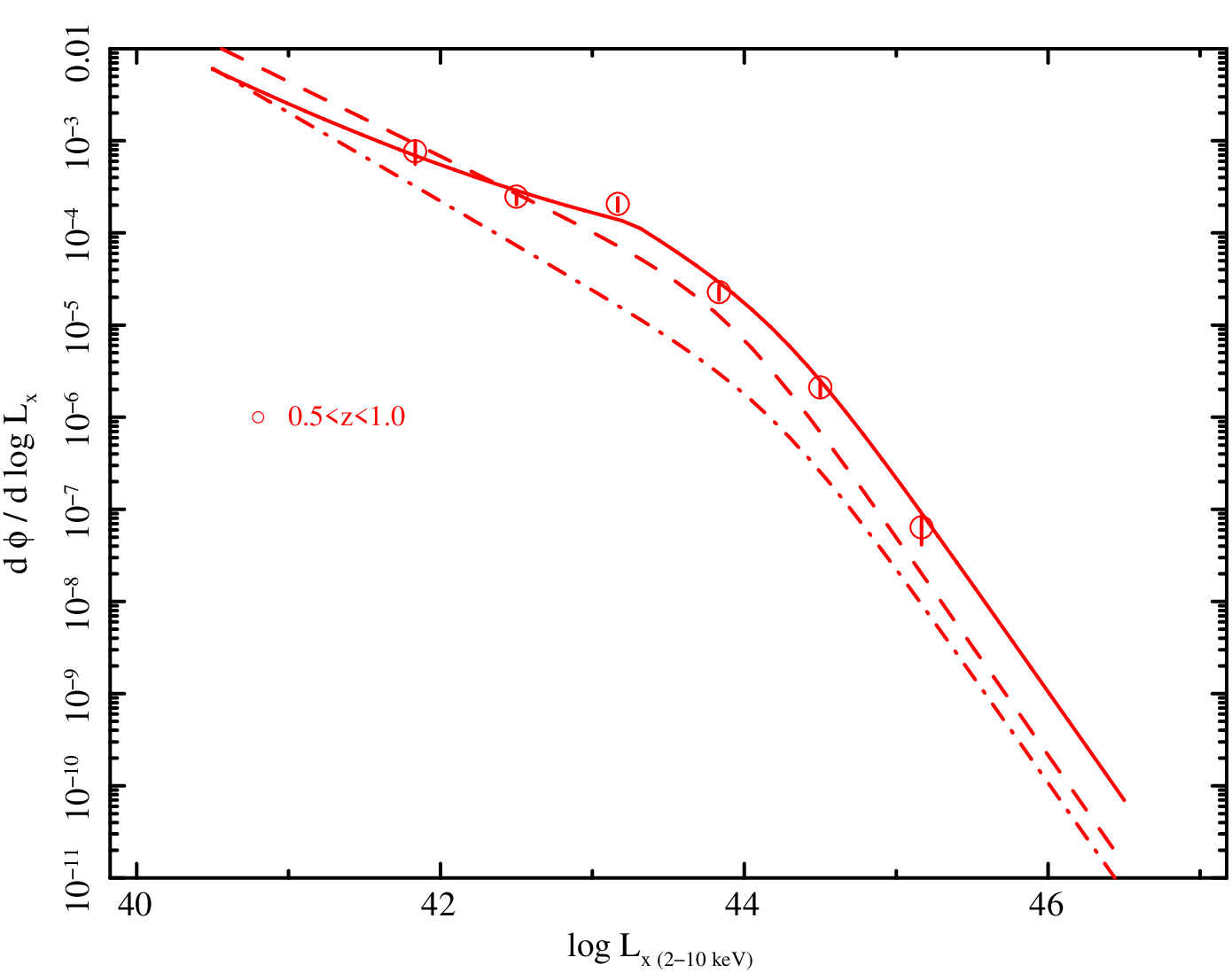}
}
\hbox{
\includegraphics[width=6.5cm,angle=270.0]{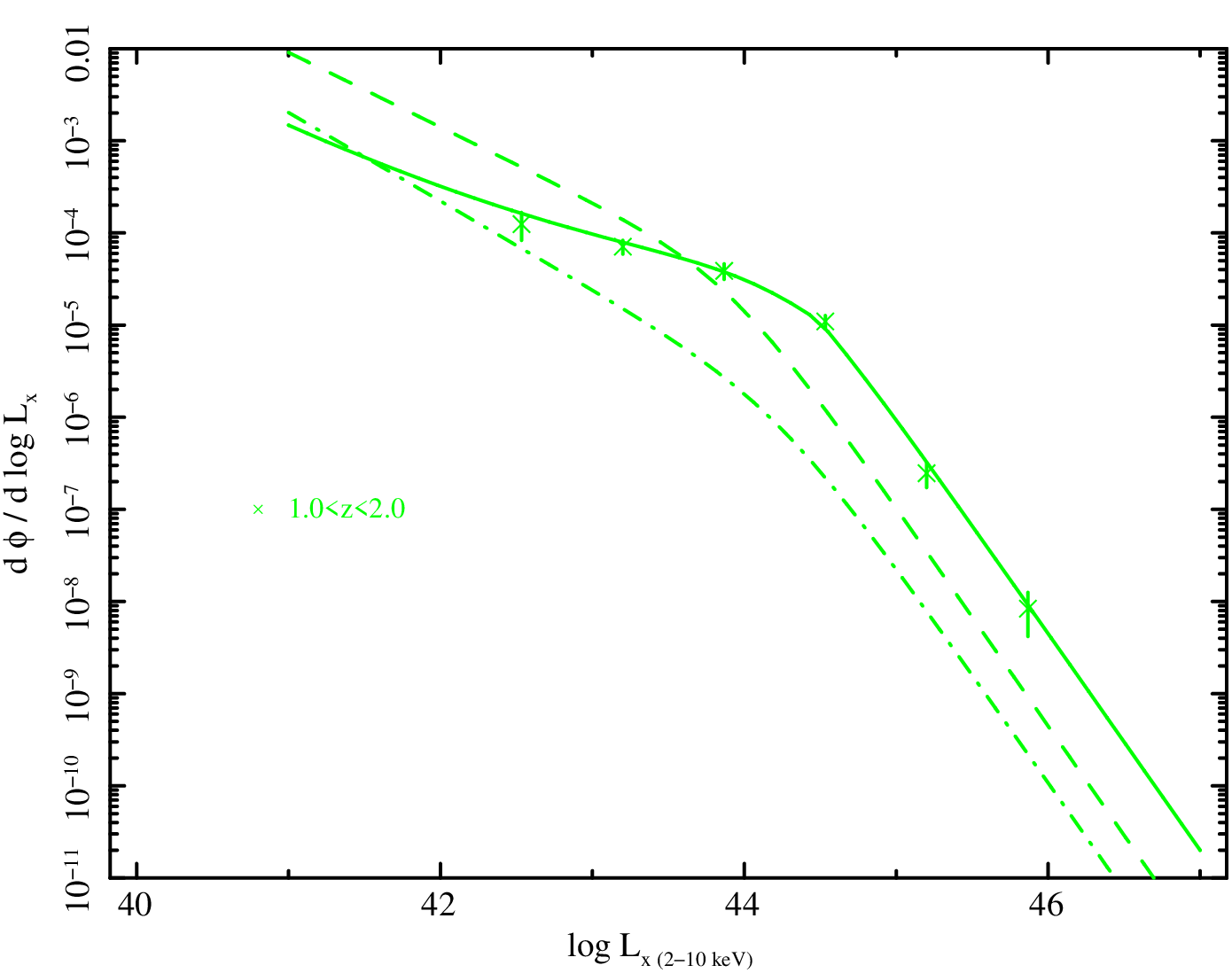}
\includegraphics[width=6.5cm,angle=270.0]{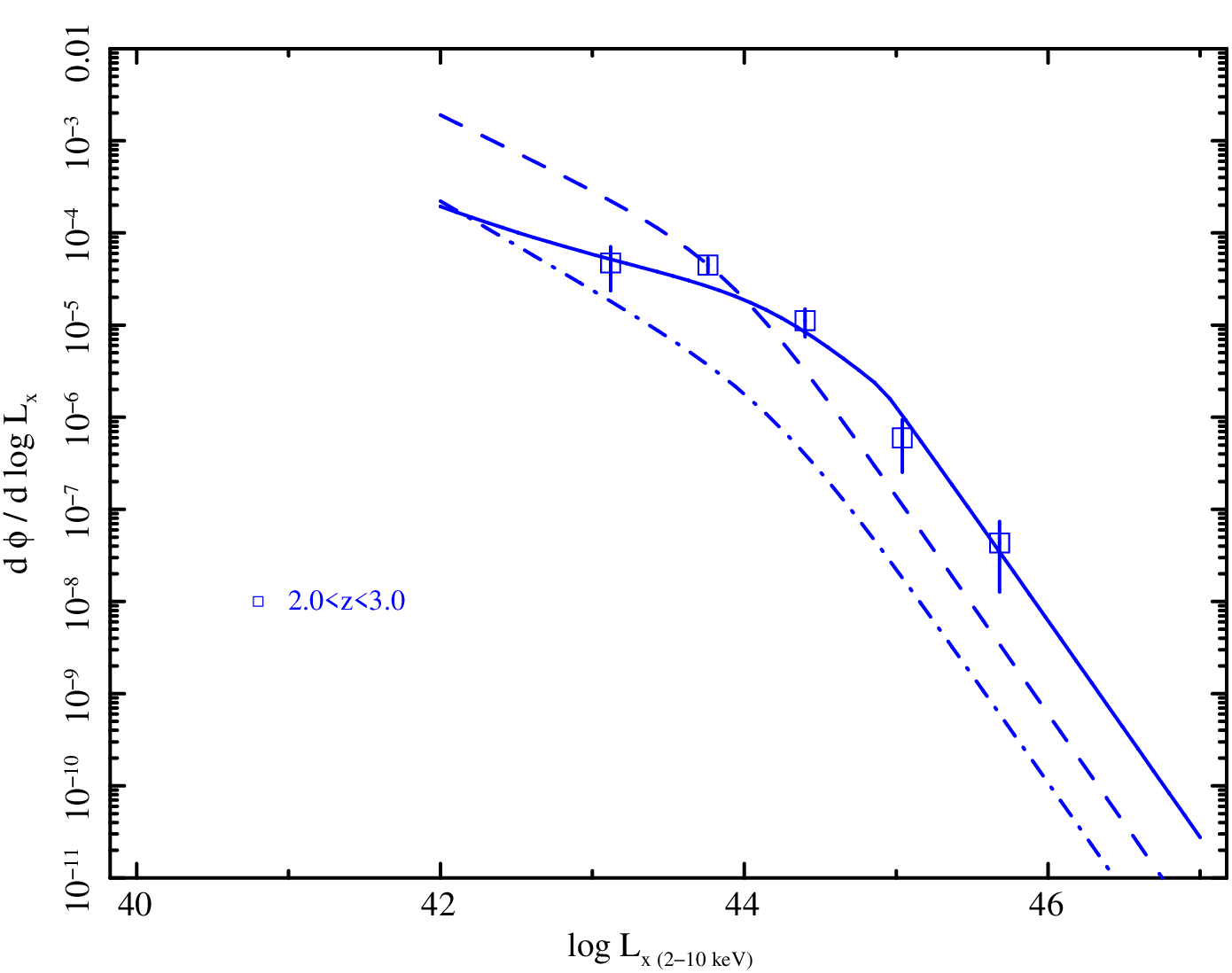}
}
\caption[]{X-ray luminosity function in the Hard band in the redshift ranges 0.01-0.5 ({\it Top left panel}), 0.5-1.0 ({\it Top right panel}), 1.0-2.0 ({\it Bottom left panel}) and 2.0-3.0 ({\it Bottom right panel}). The points are the results from the binned XLF and the lines are the best-fit curves to our analytical LDDE ({\it solid line}) and PLE ({\it dashed line}) models evaluated at the median of each redshift range. The dot-dashed line is our LDDE model evaluated at $z=0$.}
\label{fig:XLFhard}
\end{figure*}

\begin{figure*}
\centering
\hbox{
\includegraphics[width=6.5cm,angle=270.0]{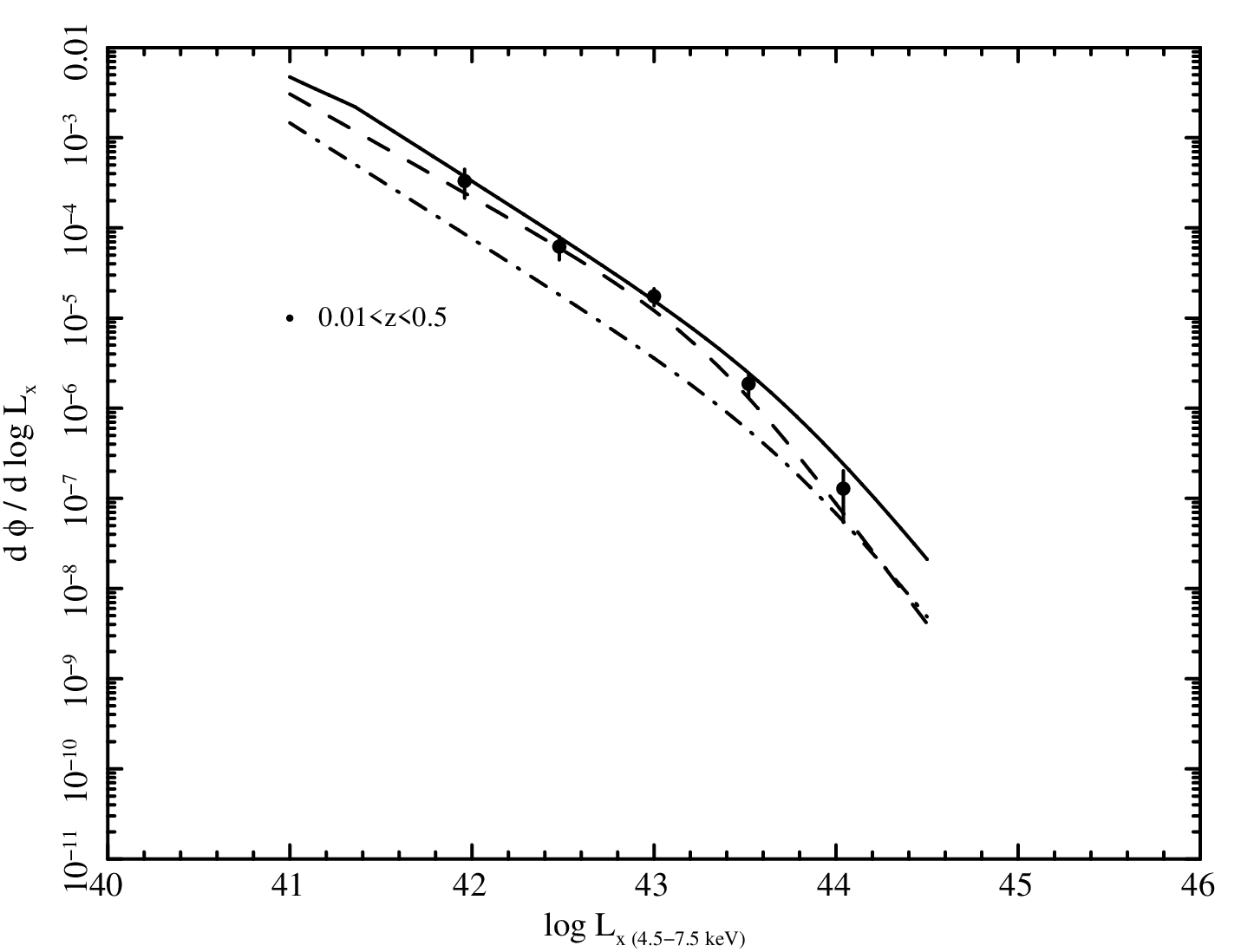}
\includegraphics[width=6.5cm,angle=270.0]{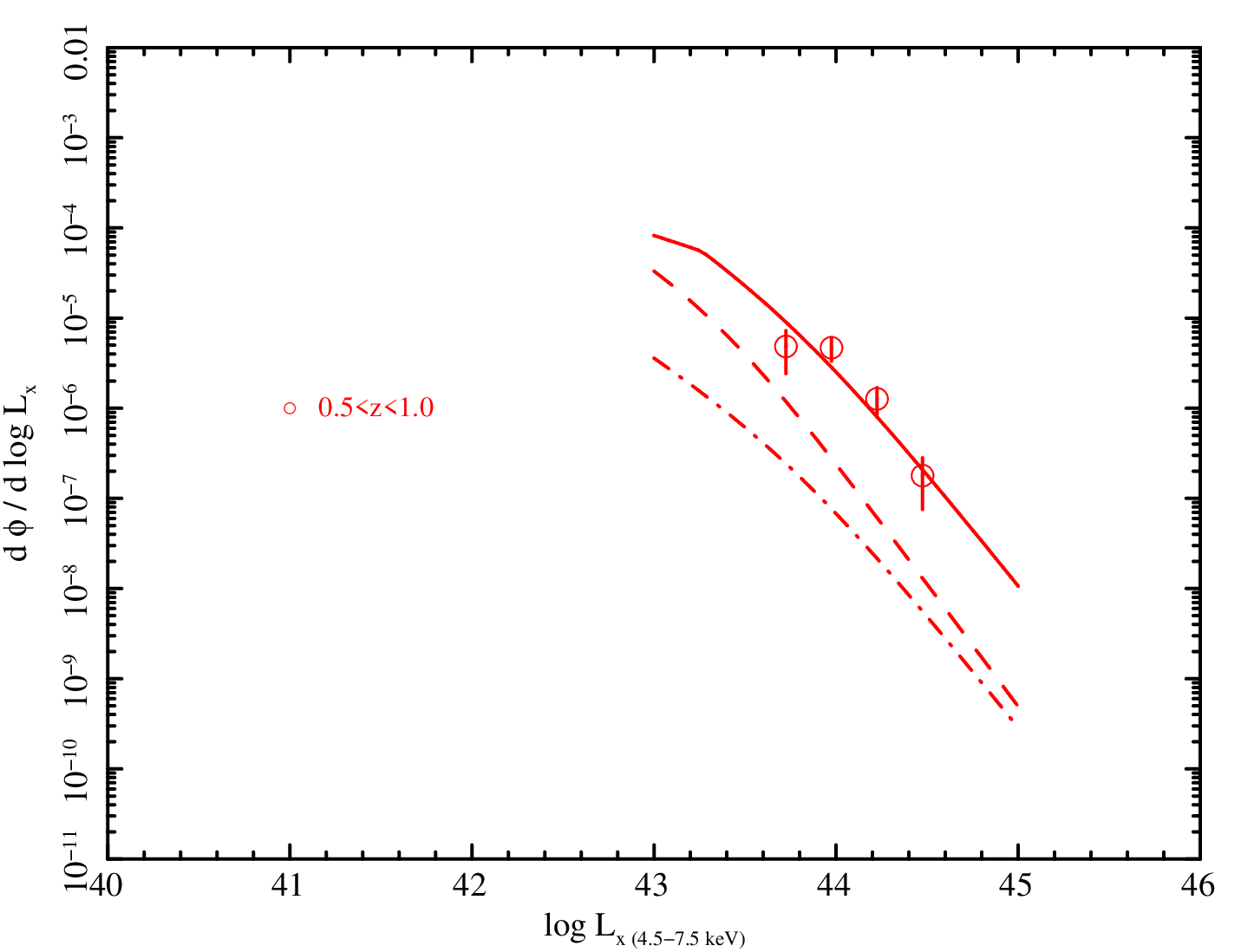}
}
\includegraphics[width=6.5cm,angle=270.0]{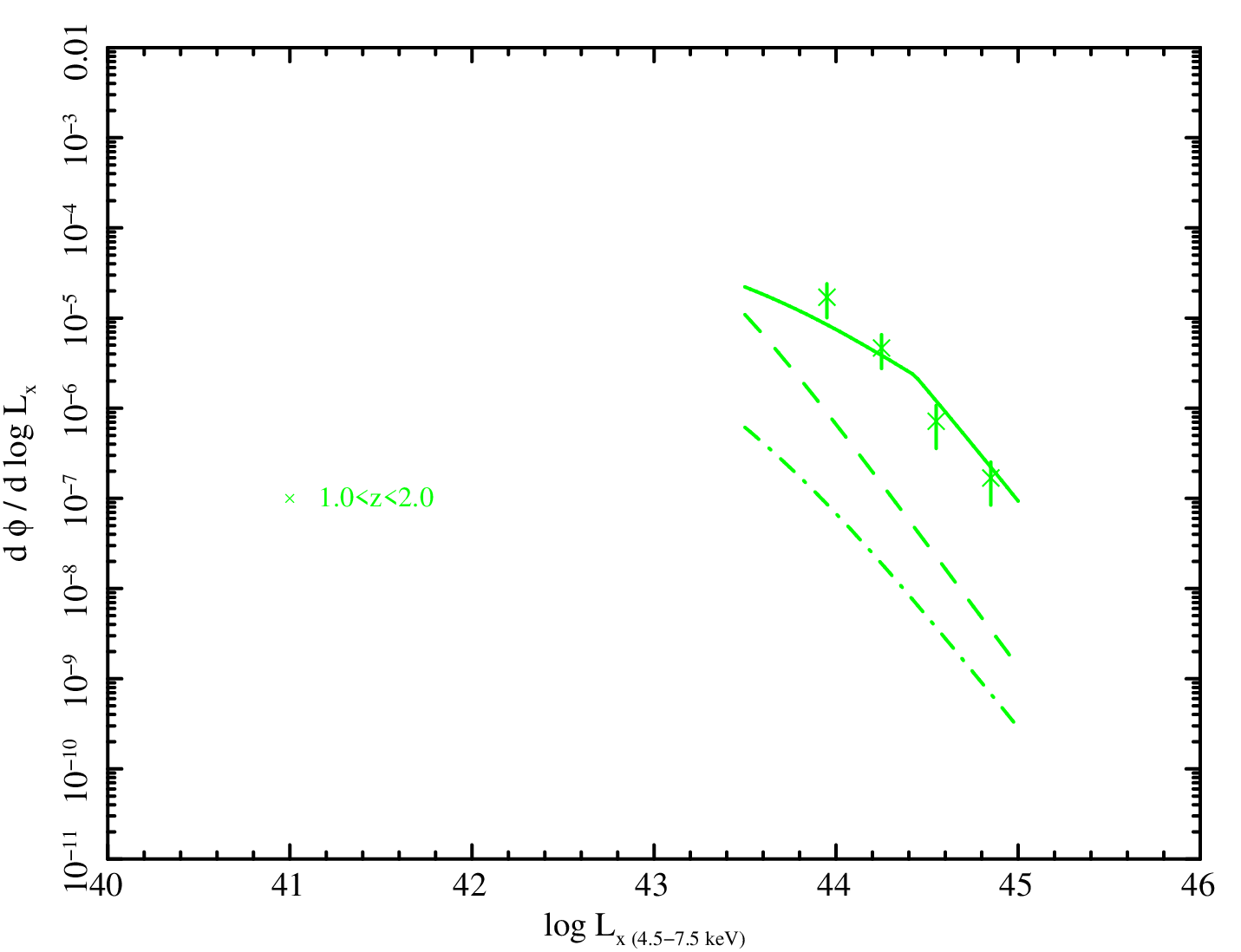}
\caption[]{X-ray luminosity function in the Ultrahard band in the redshift ranges 0.01-0.5 ({\it Top left panel}), 0.5-1.0 ({\it Top right panel}) and 1.0-2.0 ({\it Bottom panel}). The points are the results from the binned XLF and the lines are the best-fit curves to our analytical LDDE ({\it solid line}) and PLE ({\it dashed line}) models evaluated at the median of each redshift range. The dot-dashed line is our LDDE model evaluated at $z=0$.}
\label{fig:XLFuh}
\end{figure*}

The best-fit parameters of the XLF calculated by the ML method explained above are summarized in Table~\ref{tab:XLF}. Best fit results to the PLE and LDDE models along with data points from the binned XLF are shown in Figures~\ref{fig:XLFsoft}, \ref{fig:XLFhard} and \ref{fig:XLFuh}.

At first sight our best-fit models in the Soft band (0.5-2~keV) reveal strong differences between the PLE and LDDE models. Although the PLE best-fit seems to be a fair description of the XLF at low redshifts ($z<1$), it clearly fails to reproduce the behaviour at high redshifts, where the LDDE model matches better with our binned XLF data points.
 
The best-fit parameters obtained by our ML technique slightly differ from the soft XLF models reported in Miyaji et al. (\cite{Miyaji00}) and Hasinger et al. (\cite{Hasinger05}) using a variety of \ROSAT{}, \XMM{} and \Chan{} surveys. For instance, our best-fit PLE model requires less evolution ($p_1$ parameter) than that of Hasinger et al. (\cite{Hasinger05}) model and the present-day luminosity break $L_0$ is displaced to fainter luminosities in our model (but consistent within 2$\sigma$ with that of Hasinger et al.). In the Hard band, our best-fit PLE model is consistent within the 1$\sigma$ error bars with the results obtained by Barger et al. (\cite{Barger05}) in the \Chan{} Deep Field.

It has been shown in different works that the LDDE model for the XLF provides the best framework that describes the evolutionary properties of AGN, both in soft (Miyaji et al. \cite{Miyaji00}, Hasinger et al. \cite{Hasinger05}) and hard X-rays (Ueda et al. \cite{Ueda03}, La Franca et al. \cite{LaFranca05}), as well as in the optical range (Bongiorno et al. \cite{Bongiorno07}).

The LDDE model provides a better overall description of the XLF. However, our present day XLF power law shape has flatter slopes and the luminosity break $L_0$ is also displaced to fainter luminosities (similarly as in the PLE model) than in Hasinger et al. \cite{Hasinger05} by more than 2$\sigma$, although it is consistent with the LDDE models of Miyaji et al. (\cite{Miyaji00}) within the error bars. Our best-fit evolution parameter $p_1$ reveals less evolution below the cut-off redshift (fixed to the value obtained by Hasinger et al. \cite{Hasinger05}, $z_c=1.42$) and a much weaker dependence of $z_c$ on luminosity $\alpha$ than in these works. The significative less evolution $p_{1}$ found in this work with respect that of Hasinger et al. (\cite{Hasinger05}) is somewhat surprising since both samples have comparable sizes. The differences are significative even if we repeat the fits putting the fixed parameters from Hasinger et al. (\cite{Hasinger05}) at their $\pm$1$\sigma$ values, thus obtaining $p_1$ best fits ranging from $\sim$3.4 to $\sim$3.9. However, we must note that the slopes $p_1$ and $p_2$ in the Hasinger et al. LDDE model are dependent on the X-ray luminosity while ours are not. For instance in the $\log L_X=42-46$ interval, $p_1$ in the Hasinger et al. model ranges between $\sim$3.3 and $\sim$6.1. Since our XLF is extended towards lower luminosities than that of Hasinger et al., an overall lower $p_1$ parameter is consistent with a X-ray luminosity dependency.

Similarly, in the Hard band (2-10~keV) the LDDE model outperforms the PLE model at $z>0.5$, failing the latter to describe the evolution of the binned data points at the faint end of the XLF underestimating the data at $\log L_X>44$ by a factor of several. Comparing our results in the Hard band (2-10~keV) with previous major works in this band (Ueda et al. \cite{Ueda03}, La Franca et al. \cite{LaFranca05}, Silverman et al. \cite{Silverman08}) we find an overall good agreement between our work and theirs, although with few differences. The general shape of the present day XLF, described by the smooth double power law, is similar to that reported in other works within the error bars. The evolution below the redshift cut-off $p_1$ is also consistent with the results of Ueda et al. (\cite{Ueda03}) and La Franca et al. (\cite{LaFranca05}) (both using samples corrected by the intrinsic absorption of their sources), and Silverman et al. (\cite{Silverman08}) well within the 1$\sigma$ confidence level. The stronger difference between these models arises with the strength of the dependence of the cut-off redshift $z_c$ on luminosity, measured by the parameter $\alpha$. The values of $\alpha$ obtained by Ueda et al. (\cite{Ueda03}) and Silverman et al. (\cite{Silverman08}) are consistent with each other while the ones calculated by La Franca et al (\cite{LaFranca05}) ($\alpha \sim 0.2$) and in this work ($\alpha \sim 0.25$) are sistematically lower. Overall, our best-fit LDDE parameters are more constrained than in the other works according to the computed 1$\sigma$ error bars. We have left the evolution parameter above the cut-off redshift $p_2$ fixed to that of Ueda et al. (the value obtained in La Franca et al. reveals less evolution beyond $z_c$ but consistent with Ueda et al. within 1$\sigma$). The model in Silverman et al. (\cite{Silverman08}) requires a much stronger evolution $p_2=-3.27_{-0.34}^{+0.31}$ than the others. It must be noted, nevertheless, that it is extremely difficult to properly constrain the faint end of the luminosity function given the necessity of highly complete deep pencil-beam surveys that account for the population of high-redshift low-luminosity AGN.

In a recent work, Della Ceca et al. (\cite{DellaCeca08}) have computed the intrinsic present day XLF of the HBSS sample in the 4.5-7.5~keV (Ultrahard) band for absorbed and unabsorbed AGN. In this paper we have added the HBSS sample to ours, and performed the same analysis as in the Soft and Hard bands using both absorbed and unabsorbed sources. This way we have doubled the number of available AGN and improved the $L_X-z$ coverage, which is directly reflected in that the best-fit parameters are more constrained and the associated error bars are significantly smaller. Moreover, given the availability of detailed spectral information for the vast majority of the sources in both samples, we have carried out the XLF analysis by convolving the $N_H$ function with the XLF analytical model when performing the fit (see section~\ref{XLFuh}). Like in the other energy bands, the PLE best-fit clearly underestimates the data at faint luminosities and high redshifts. From the binned data points it can be inferred that a very strong evolution $p_1$ is required to describe properly the behaviour of the Ultrahard sources, which is achieved by the LDDE model. Our results are fully consistent with those reported in Della Ceca et al. (\cite{DellaCeca08}) albeit with smaller error bars. The shape of our best-fit present day XLF would correspond to the average of the absorbed and unabsorbed present day XLF calculated in Della Ceca et al. thus lying inside the 1$\sigma$ confidendence levels spanned by their parameters. Our best-fit value for the evolution parameter $p_1=6.46_{-0.29}^{+0.69}$ is in excellent agreement with that of Della Ceca et al. (\cite{DellaCeca08}) ($p_1=6.5$) and also with that of Bongiorno et al. (\cite{Bongiorno07}) obtained from a selected AGN sample in the optical range from the VIMOS-VLT survey ($p_1=6.54$) and that of the bolometric quasar luminosity function of Hopkins et al. (\cite{Hopkins07}) ($p_1=5.95\pm0.23$). This value represents a much stronger evolution below the cut-off redshift than that obtained by Ueda et al. (\cite{Ueda03}) ($p_1=4.23\pm0.39$) and La Franca et al. (\cite{LaFranca05}) ($p_1=4.62\pm0.26$) in the 2-10~keV band.

\subsection{Accretion history of the Universe}
\label{accretion}

\begin{figure}
\centering
\hbox{
\includegraphics[width=6.5cm,angle=270.0]{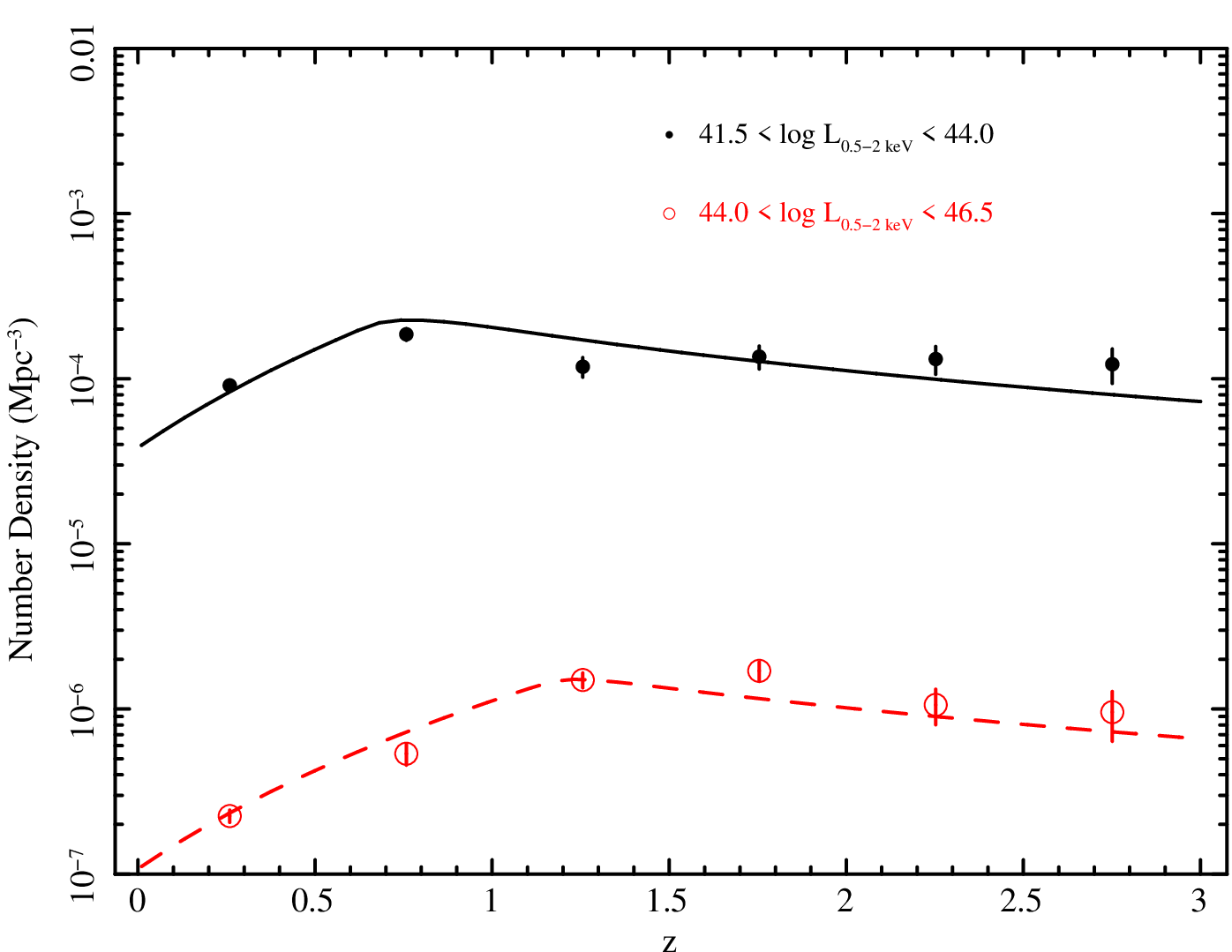}
}
\hbox{
\includegraphics[width=6.5cm,angle=270.0]{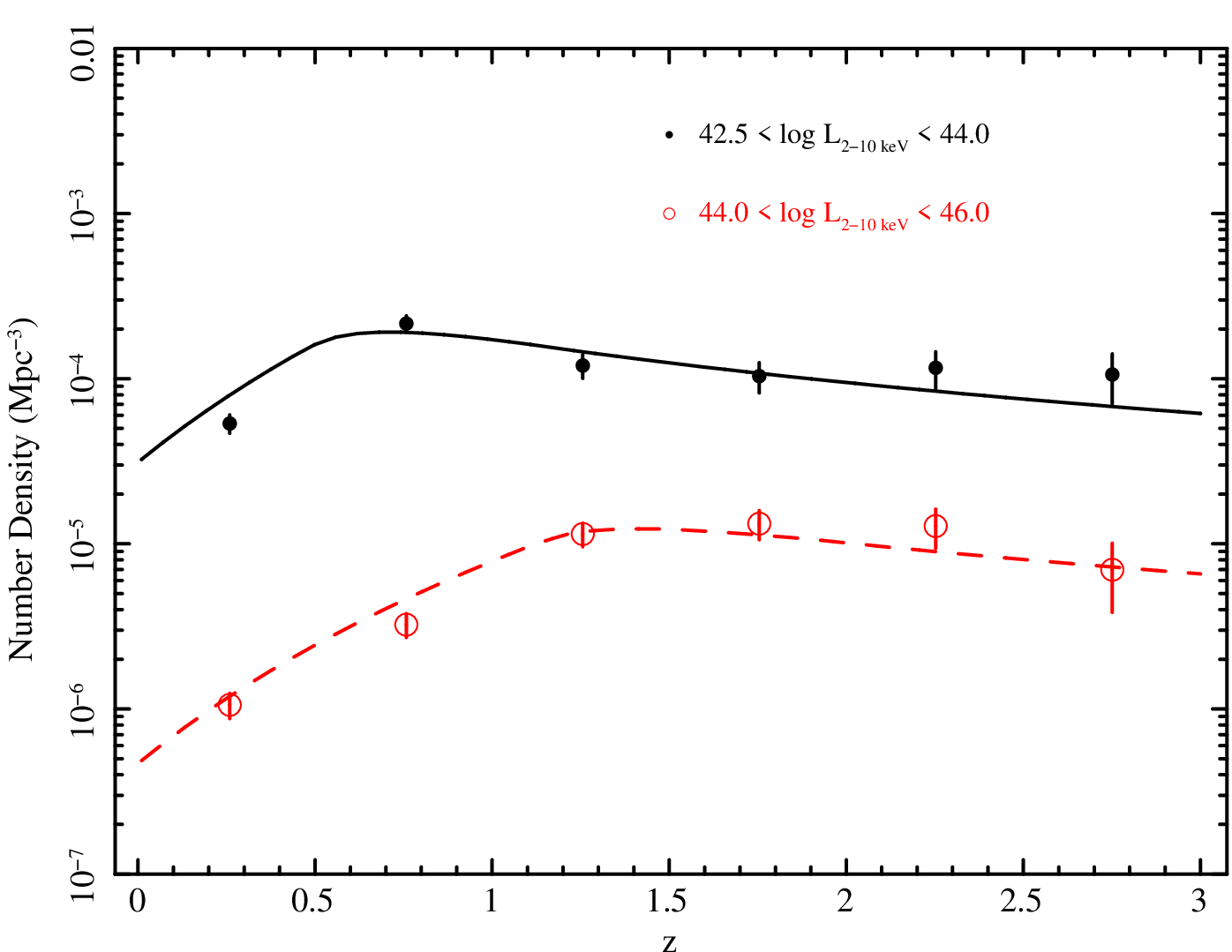}
}
\hbox{
\includegraphics[width=6.5cm,angle=270.0]{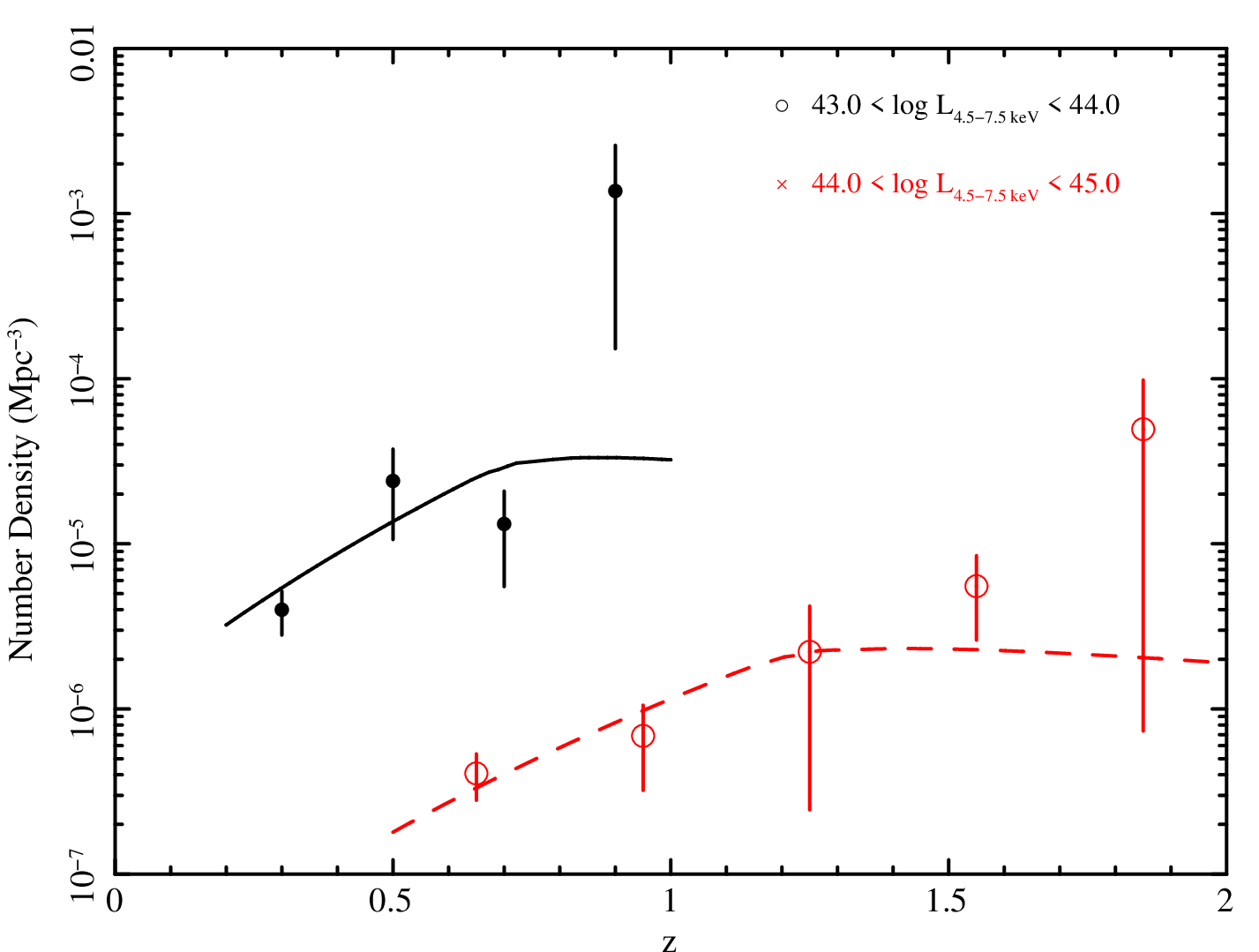}
}
\caption[]{Comoving density of AGN in different luminosity bins as a function of redshift in the Soft ({\it top panel}), Hard ({\it Center panel}) and Ultrahard ({\it Bottom panel}) bands. Overplotted are the predictions from our best-fit LDDE model.}
\label{fig:ndensity}
\end{figure}

The X-ray luminosity function of AGN, whose bolometric emission is directly linked to accretion power, constrains the history of the formation of the supermassive black holes that reside in the galactic centers along cosmic times.

The comoving density of AGN as a function of redshift can be calculated straightforward from our best-fit XLF LDDE model:

\begin{equation}
\label{eq:ndensity}
\Phi(z)=\int_{L_1}^{L_2}\frac{d\Phi(L_X,z)}{d\log L_X}d\log L_X
\end{equation}

As we can see in Figure~\ref{fig:ndensity}, the most luminous AGN are formed before the less luminous ones in all bands. AGN with $\log L_X > 44$ reach a maximum in density at redshift $\sim$1.5 while fainter AGN ($\log L_X < 44$) peak at $z \sim 0.7$.

Similarly to the comoving density, we can derive the luminosity density as a function of redshift in all bands and, from that, calculate the accretion rate density. Here we will assume that the accretion rate onto a supermassive black hole is related to the bolometric luminosity by a constant factor $\epsilon$ which is the radiative efficiency of the accretion flow (Marconi et al. \cite{Marconi04}, La Franca et al. \cite{LaFranca05}):

\begin{equation}
\label{eq:Macc}
L_{bol}=\frac{\epsilon}{1-\epsilon}\dot{M}_{acc}c^2
\end{equation}

We can derive the bolometric luminosities by means of a bolometric correction factor $K$ simply using $L_{bol}=K L_X$. The accretion rate density is hence (Soltan \cite{Soltan82}, Marconi et al. \cite{Marconi04}):

\begin{equation}
\label{eq:accdensity}
\dot{\rho}_{acc}(z)=\frac{1-\epsilon}{\epsilon c^2}\int_{L_1}^{L_2}K L_X \frac{d\Phi(L_X,z)}{d\log L_X} d\log L_X
\end{equation}

\noindent We will assume a nominal radiative efficiency of $\epsilon=0.1$ (Yu \& Tremaine \cite{Yu02}, Marconi et al. \cite{Marconi04}, Barger et al. \cite{Barger05}). The values of the bolometric correction $K$ are derived from the polynomic expressions of Marconi et al. (\cite{Marconi04}), that accounts for changes in the overall spectral energy distribution of AGN as a function of the optical luminosity. Hence, the total accreted mass onto supermassive black holes is:

\begin{equation}
\label{eq:massdensity}
\rho(z)=\int_{z}^{z_0}\dot{\rho}_{acc}(z)\frac{dt}{dz}dz
\end{equation}

\noindent where we assume that the initial mass of seed black holes at $z_0$ is negligible with respect to the total accreted mass (La Franca et al. \cite{LaFranca05}).

\begin{figure}
\centering
\hbox{
\includegraphics[width=6.5cm,angle=270.0]{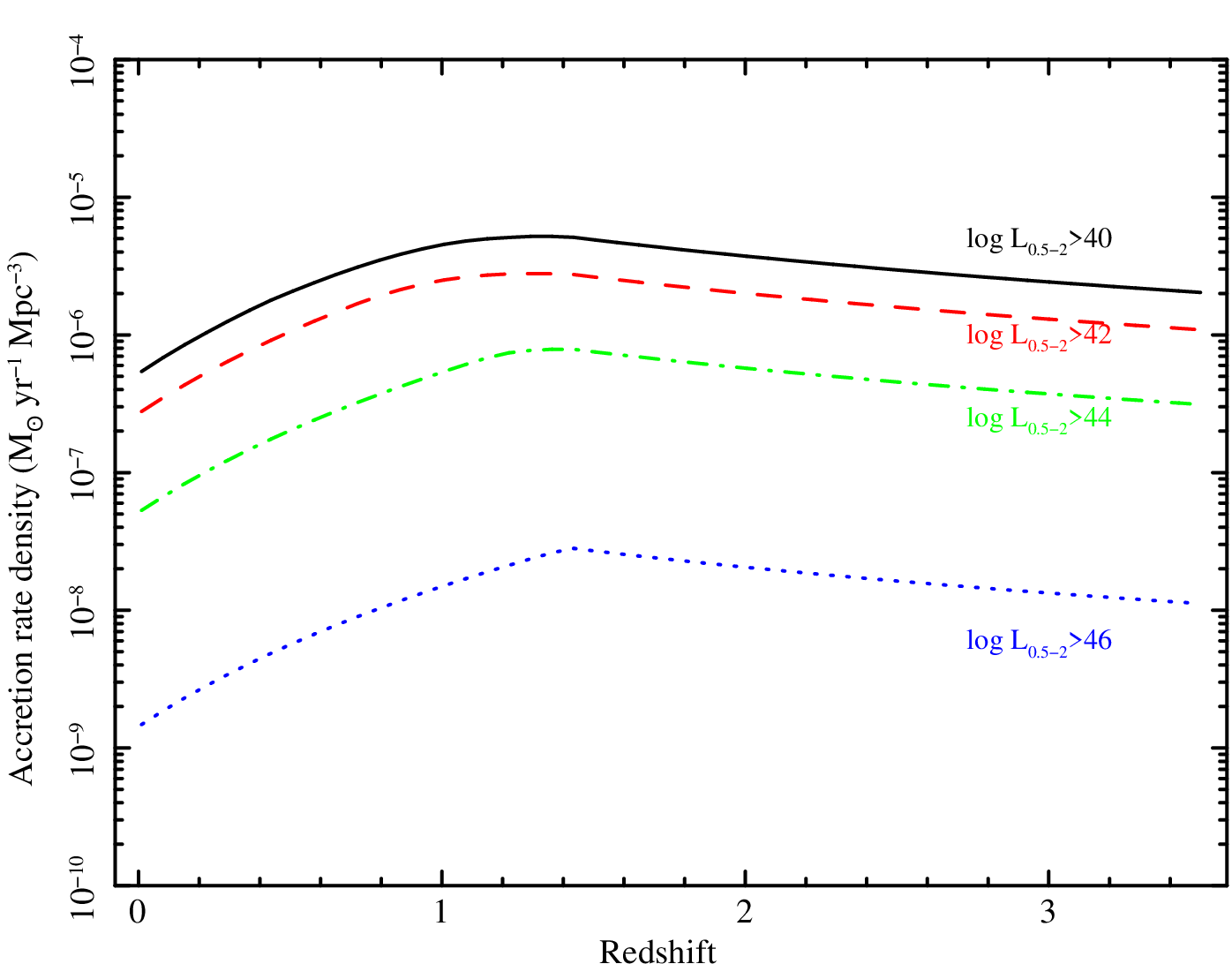}
}
\hbox{
\includegraphics[width=6.5cm,angle=270.0]{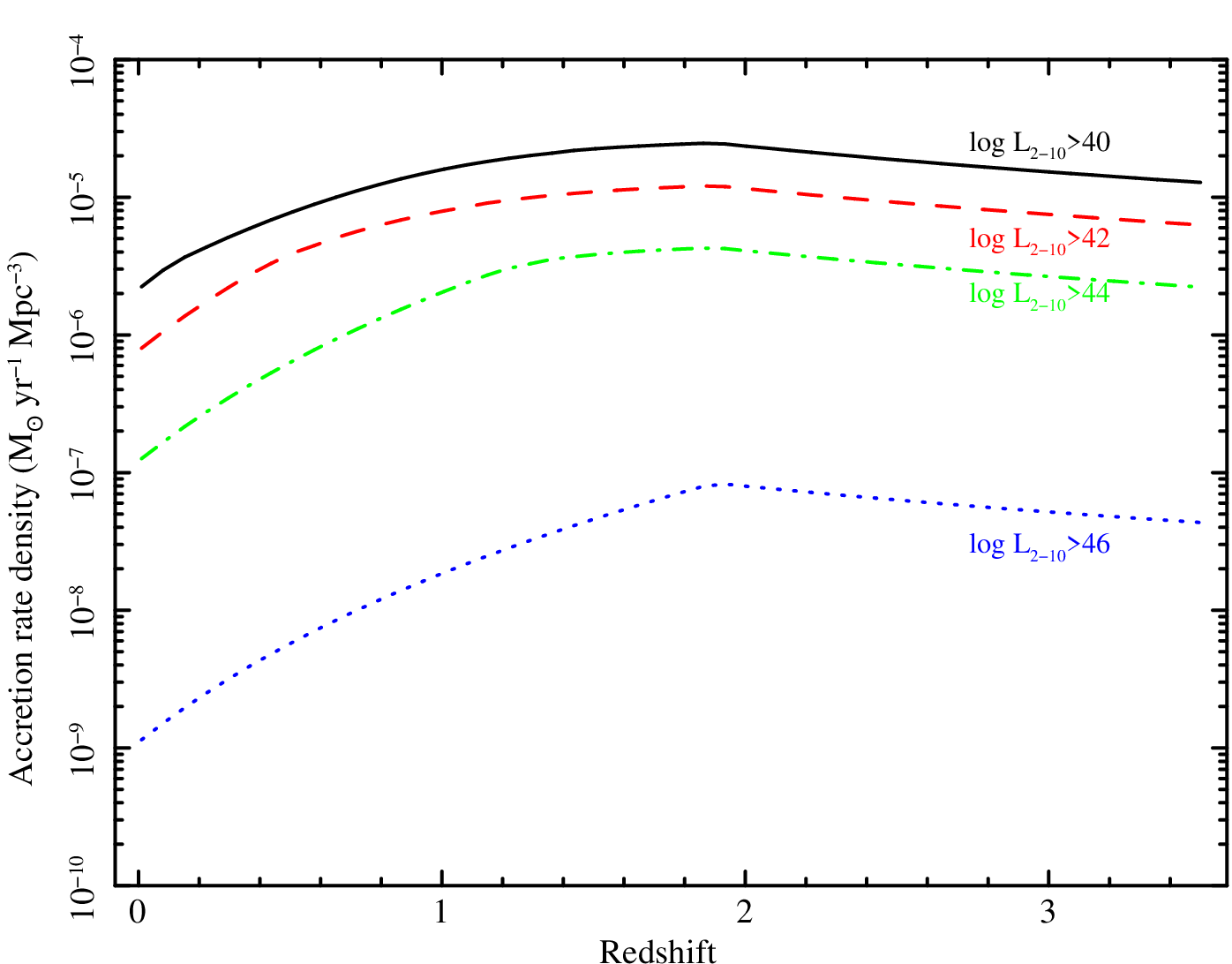}
}
\hbox{
\includegraphics[width=6.5cm,angle=270.0]{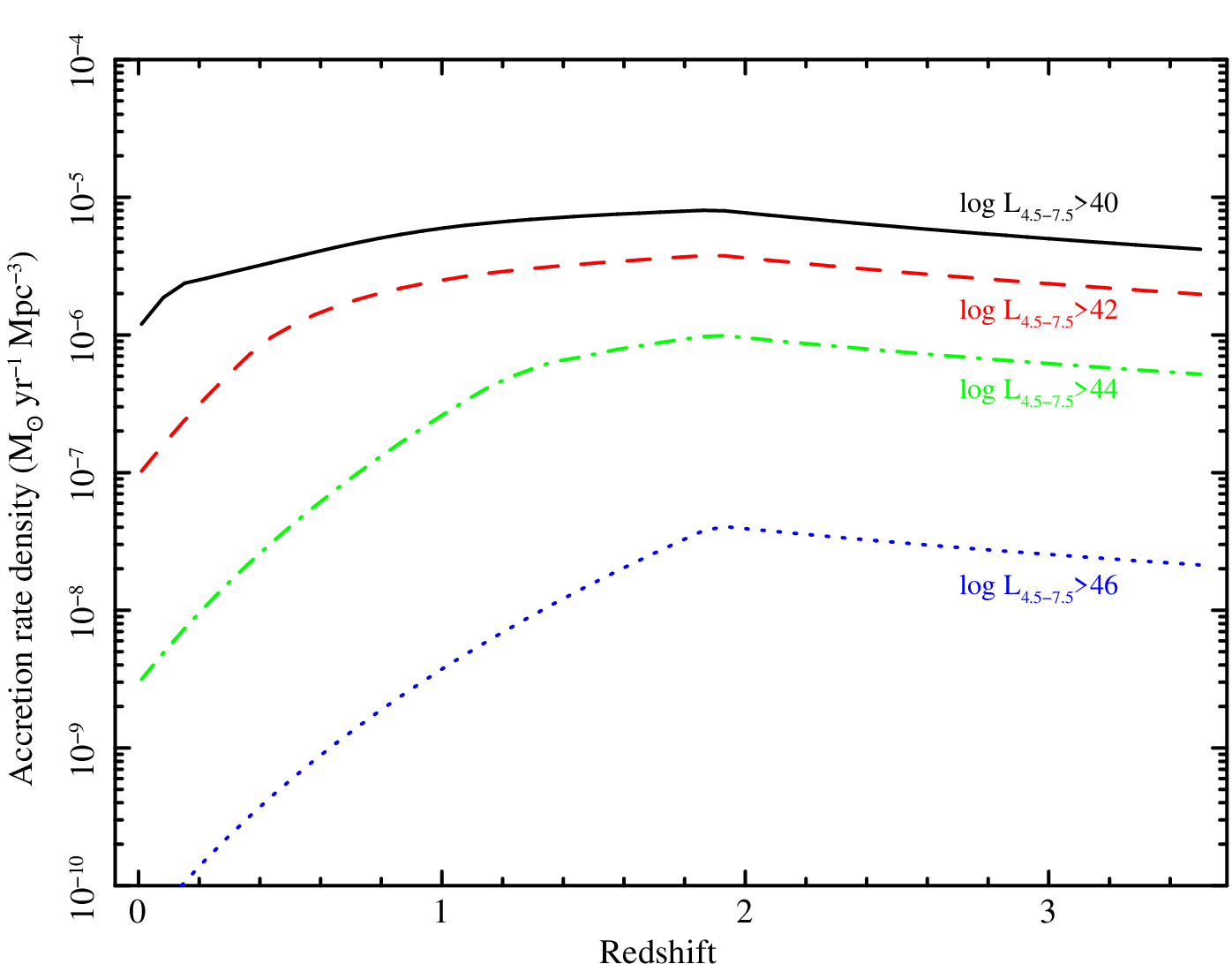}
}
\caption[]{Accretion rate density of matter onto supermassive black holes as a function of redshift in the Soft ({\it top panel}), Hard ({\it Center panel}) and Ultrahard ({\it Bottom panel}) bands.}
\label{fig:accdensity}
\end{figure}

\begin{figure}
\centering
\hbox{
\includegraphics[width=6.5cm,angle=270.0]{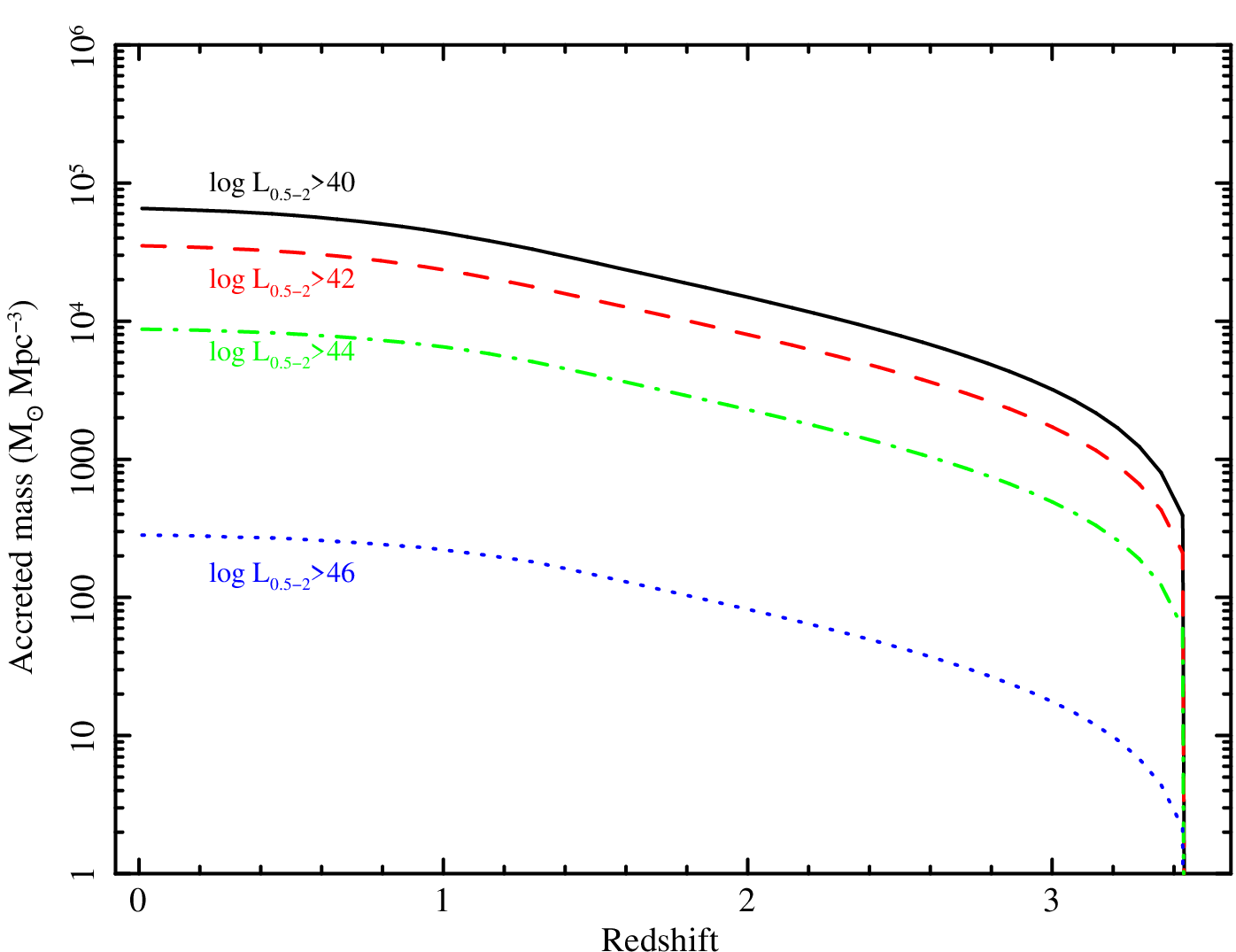}
}
\hbox{
\includegraphics[width=6.5cm,angle=270.0]{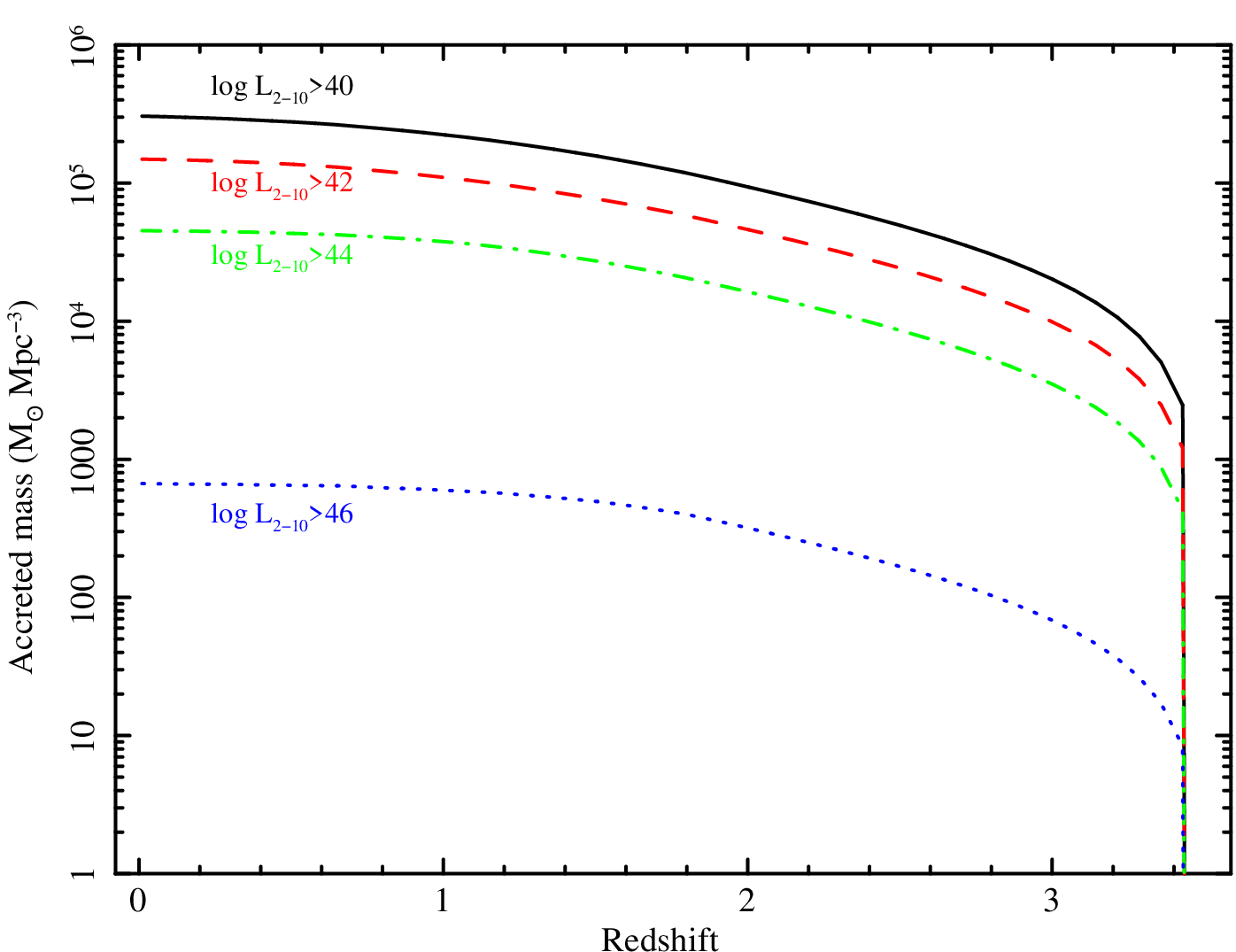}
}
\hbox{
\includegraphics[width=6.5cm,angle=270.0]{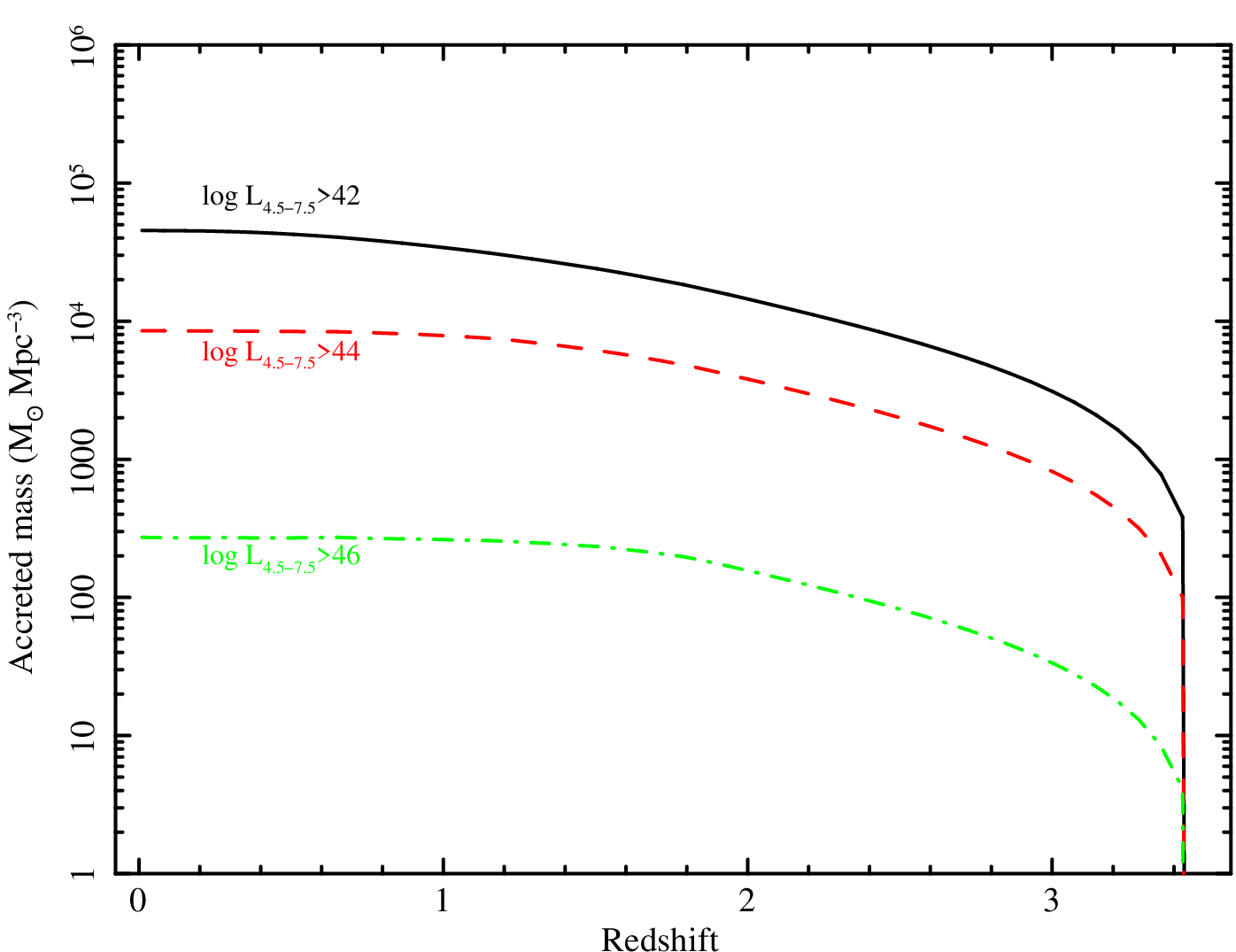}
}
\caption[]{Total accreted mass onto supermasive black holes as a function of redshift in the Soft ({\it top panel}), Hard ({\it Center panel}) and Ultrahard ({\it Bottom panel}) bands.}
\label{fig:massdensity}
\end{figure}

In Figures~\ref{fig:accdensity} and \ref{fig:massdensity} is clearly seen that the vast majority of the accretion rate density and the total mass density are produced by low-luminosity AGN ($\log L_X<44$). The total accreted mass density at $z=0$ obtained from our XLF model ($\sim 3\times 10^5$~M$_\odot$ Mpc$^{-3}$) in the hard band is in good agreement with that of La Franca et al. (\cite{LaFranca05})($3.2 h_{70}^{2} \times 10^5$~M$_\odot$ Mpc$^{-3}$) derived from their intrinsic XLF, or the local black hole mass function of Marconi et al. (\cite{Marconi04}) ($4.6_{-1.4}^{+1.9}h_{70}^{2}\times 10^5$~M$_\odot$ Mpc$^{-3}$) determined applying the correlations between black hole mass, bulge luminosity and stellar luminosity dispersion to galaxy luminosity and velocity functions in the Local Universe. 

From these results it can be inferred that high-luminosity AGN grow and feed very efficiently in the early Universe and are fully formed at redshifts 1.5-2, whereas the low-luminosity AGN keep forming down to $z\sim 1$. These results are hence in full agreement with an anti-hierarchical black hole growth scenario as shown by the LDDE model of the XLF in a number of previous works (e.g. Ueda et al. \cite{Ueda03}, Merloni \cite{Merloni04}, Marconi et al. \cite{Marconi04}, La Franca et al. \cite{LaFranca05}, Hasinger et al. \cite{Hasinger05}).


\section{Conclusions}
\label{conclusions}

We have discussed here the cosmic evolution of a sample of AGN in three X-ray bands: Soft (0.5-2~keV), Hard (2-10~keV) and Ultrahard (4.5-7.5~keV). The backbone of our sample is the XMS survey (Barcons et al. \cite{Barcons07}) which is a flux-limited highly-complete sample at medium fluxes. We have combined the XMS with other shallower and deeper highly complete X-ray surveys in all three bands to end up with a total sample of $\sim$1000, 435 and 119 AGN in the Soft, Hard and Ultrahard bands, respectively.

 We have used the spectral information on the sources that compose the Hard (2-10~keV) band to model their instrinsic absorption ($N_H$ function). We find dependency of the fraction of absorbed AGN on both the X-ray luminosity and redshift. Our predictions on the behaviour of the fraction of absorbed AGN in this band is in excellent agreement with the results of Ueda et al. (\cite{Ueda03}) and La Franca et al. (\cite{LaFranca05}). In recent work, Hasinger (\cite{Hasinger08}) studies the evolution of the absorption properties of AGN using a compiled sample of 1290 AGN in the 2-10~keV band, being our results in good agreement with his within 1$\sigma$. The same analysis was applied to the Ultrahard samples, XMS and HBSS (Della Ceca et al. \cite{DellaCeca08}). We find that the fraction of absorbed of AGN in the 4.5-7.5~keV band, assuming a dividing value of $N_H=10^{22}$~cm$^{-2}$, is dependent on the X-ray luminosity but not on the redshift. This could be motivated by the narrow redshift range spanned by the sample, with the bulk of the sources located at low redshifts ($z<1$). In the same band, Della Ceca et al. (\cite{DellaCeca08}) suggest a possible evolution of the absorbed AGN population with redshift by comparing their results with other survey projects.

We have calculated the X-ray luminosity function of AGN using two methods. First, a modified version of the $1/V_a$ method (Schmidt \cite{Schmidt68}) discussed in Page \& Carrera (\cite{Page00}) to compute the binned XLF. Secondly, a fit to an analytic model using a Maximum Likelihood technique (Marshall et al. \cite{Marshall83}) that fully exploits the available information on each individual source without binning. The adopted model consists of a smoothly connected double power law that accounts for the present day ($z=0$) XLF modified by an evolution factor that depends on the redshift (PLE model) or on both redshift and luminosity (LDDE model). We have found that the LDDE model outperforms the PLE model at faint luminosities and high redshifts.

In general, our computed XLF in the Soft, Hard and Ultrahard bands are in good agreement with the results of other published surveys. The best-fit XLF parameters in the Soft sample show slight discrepancies in both the overall shape and evolution with respect previous works in that band (Miyaji et al. \cite{Miyaji00}, Hasinger et al. \cite{Hasinger05}). In the Hard band, where we have computed the intrinsic XLF taking into account the intrinsic absorption $N_H$ of each source, we are in good agreement with the results of La Franca et al. (\cite{LaFranca05}) although there are differences in detail with Ueda et al. (\cite{Ueda03}) or Silverman et al. (\cite{Silverman08}). Our best-fit model shows weaker evolution of the AGN below the cut-off redshift than in these works albeit with smaller error bars. In the Ultrahard band, we have also calculated the intrinsic XLF finding similar results as in Della Ceca et al. (\cite{DellaCeca08}) but with our best-fit parameters much more constrained. The results in this band show that Ultrahard AGN present a significantly stronger evolution below the cut-off redshift than those detected at softer energies. A possible explanation for this could reside in the existence of a dependency on the X-ray luminosity for the evolutionary parameter $p_1$ that we have not taken into account in our model due to the low statistics in this band. Such a dependency has been found in the XLF analysis in other bands (e.g. Ueda et al. \cite{Ueda03}, Hasinger et al. \cite{Hasinger05}) showing that $p_1$ linearly increases with luminosity. Given that in the Ultrahard band we lack of deep surveys, which account mainly for intrinsically faint objects, our result could be biased towards high-luminosity sources hence providing higher evolution rates below the cut-off redshift. In all three bands, the high-luminosity AGN ($\log L_X > 44$) are formed before than the low-luminosity ones ($\log L_X < 44$), reaching the former a maximum in density at redshift $z\sim1.5$ whereas the comoving density of the latter peak at $z\sim 0.7$.

Finally, we have used our best-fit XLF to compute the accretion rate density and total accreted mass onto supermassive black holes as a function of redshift. The total black hole mass density at $z=0$ predicted by our best-fit model is in agreement with that computed by La Franca et al. (\cite{LaFranca05}) and the local supermassive black hole density derived by Marconi et al. (\cite{Marconi04}). Although the value derived by Marconi et al. (\cite{Marconi04}) is slightly higher than those computed by La Franca et al. (\cite{LaFranca05}) and in this work, it has to be taken into account that hard surveys like the ones used here still miss the Compton-thick AGN population which are contributors to the relic black hole mass function. Marconi et al. (\cite{Marconi04}) estimate the contribution of Compton-thick AGN to the observed BH mass function in a factor of $\sim$1.5 which would fully explain this discrepancy. Similarly, Della Ceca et al. (\cite{DellaCeca08}) estimate a density ratio between Compton-thick and Compton-thin AGN of 1.08$\pm$0.44 at $\log L_X \sim 43$.

As predicted by the XLF LDDE model, the high-luminosity AGN have a more efficient growth in the early stages of the Universe and are fully formed at $z\sim 1.5-2$ while the less luminous AGN keep forming down to redshifts below 1 (see e.g. Merloni \cite{Merloni04}, Marconi et al. \cite{Marconi04}, Hasinger et al. \cite{Hasinger05} among others). This behaviour is found in all energy bands under study, thus confirming that the evolution of the intrinsic XLF along cosmic time is not caused by changes in the absorption environment but by intrinsic variations in the accretion rate at different epochs of the Universe.


\begin{acknowledgements}
The authors wish to thank Dr. Takamitsu Miyaji for useful discussion. This research has made use of the NASA/IPAC Extragalactic Database (NED; which is operated by the Jet Propulsion Laboratory, California Institute of Technology, under contract with the National Aeronautics and Space Administration) and of the SIMBAD database (operated by CDS, Strasbourg, France). Financial support for this work was provided by the Spanish Ministerio de Educaci\'on y Ciencia under project ESP2006-13608-C02-01. RDC acknowledges financial support from the MIUR (grant PRIN-MIUR 2006-02-5203) and from the Italian Space Agency (ASI, grants m. I/088/06/0). We thank the anonymous referee for useful comments that significantly improved this paper.
\end{acknowledgements}



\begin{thebibliography}{}

\bibitem[2003]{Akiyama03} Akiyama M., Ueda Y., Ohta K., et al., 2003, ApJS, 148, 275
\bibitem[2006]{Akylas06} Akylas A., Georgantopoulos I., Georgakakis A., et al., 2006, A\&A, 459, 693
\bibitem[2003]{Alexander03} Alexander D.M., Bauer F.E., Brandt W.N., et al., 2003, AJ, 126, 539
\bibitem[1980]{Avni80} Avni Y. \& Bahcall J.N., 1980, ApJ, 235, 694
\bibitem[2002]{Baldi02} Baldi A., Molendi S., Comastri A., et al., 2002, ApJ, 564, 190
\bibitem[2006]{Ballantyne06} Ballantyne D.R., Everett J.E. \& Murray N., 2006, ApJ, 639, 740
\bibitem[2007]{Barcons07} Barcons X., Carrera F.J., Ceballos M.T., et al., 2007, A\&A, 476, 1191
\bibitem[2005]{Barger05} Barger A.J., Cowie L.L., Mushotzky R.F., et al., 2005, AJ, 129, 578
\bibitem[2004]{Bauer04} Bauer F.E., Alexander D.M., Brandt W.N., et al., 2004, AJ, 128, 2048
\bibitem[2000]{Benitez00} Ben\'itez, N., 2000, ApJ, 536, 571
\bibitem[2000]{Bolzonella00} Bolzonella M., Miralles J.M. \& Pell\'o R., 2000, A\&A, 363, 476
\bibitem[2007]{Bongiorno07} Bongiorno A., Zamorani G., Gavignaud I., et al., 2007, A\&A, 472, 443
\bibitem[1993]{Boyle93} Boyle B.J., Griffiths R.E., Shanks T., et al., 1993, MNRAS, 260, 49
\bibitem[2008]{Caccianiga08} Caccianiga A., Severgnini P., Della Ceca R., et al., 2008, A\&A, 477, 735
\bibitem[1998]{Cagnoni98} Cagnoni I., Della Ceca R. \& Maccacaro T., 1998, ApJ, 493, 54
\bibitem[2007]{Carrera07} Carrera F.J., Ebrero J., Mateos S., et al., 2007, A\&A, 469, 27
\bibitem[1995]{Comastri95} Comastri A., Setti G., Zamorani G. \& Hasinger G., 1995, A\&A, 196, 1
\bibitem[2004]{DellaCeca04} Della Ceca R., Maccacaro T., Caccianiga A., et al., 2004, A\&A, 428, 383
\bibitem[1992]{DellaCeca92} Della Ceca R., Maccacaro T., Gioia I.M., et al., 1992, ApJ, 389, 491
\bibitem[2008]{DellaCeca08} Della Ceca R., Caccianiga A., Severgnini P., et al., 2008, A\&A, 487, 119
\bibitem[2006]{Dwelly06} Dwelly T. \& Page M.J., 2006, MNRAS, 372, 1755
\bibitem[1996]{Ellis96} Ellis R.S., Colless M., Broadhurst T., et al., 1996, MNRAS, 280, 235
\bibitem[1992]{Fabian92} Fabian A.C. \& Barcons X., 1992, ARA\&A, 30, 429
\bibitem[1998]{Fabian98} Fabian A.C., Barcons X., Almaini O. \& Iwasawa K., 1998, MNRAS, 297, L11
\bibitem[1999]{Fiore99} Fiore F., La Franca F., Giommi P., et al., 1999, MNRAS, 306, L55
\bibitem[1998]{Fischer98} Fischer J.U., Hasinger G., Schwope A.D., et al., 1998, AN, 6, 347
\bibitem[2001]{Giacconi01} Giacconi R., Rosati P., Tozzi P., et al., 2001, ApJ, 551, 624
\bibitem[2002]{Giacconi02} Giacconi R., Zirm A., Wang J., et al., 2002, ApJS, 139, 369
\bibitem[2007]{Gilli07} Gilli R., Comastri A. \& Hasinger G., 2007, A\&A, 463, 79
\bibitem[2003]{Harrison03} Harrison F.A., Eckart M.E., Mao P.H., et al., 2003, ApJ, 596, 944
\bibitem[2008]{Hasinger08} Hasinger G., 2008, A\&A, in press, astro-ph/0808.0260
\bibitem[1998]{Hasinger98} Hasinger G., Burg R., Giacconi R., et al., 1998, A\&A, 329, 482
\bibitem[2005]{Hasinger05} Hasinger G., Miyaji T. \& Schmidt M., 2005, A\&A, 441, 417
\bibitem[1999]{Hogg99} Hogg D.W., 1999, astro-ph/9905116
\bibitem[2007]{Hopkins07} Hopkins P.F., Richards G.T. \& Hernquist L., 2007, ApJ, 654, 731
\bibitem[1994]{James94} James F., 1994, MINUIT reference manual, CERN Program Library Writeup D506, CERN, Geneva
\bibitem[1995]{Kormendy95} Kormendy J. \& Richstone D., 1995, ARA\&A, 33, 581
\bibitem[2005]{LaFranca05} La Franca F., Fiore F., Comastri A., et al., 2005, ApJ, 635, 864
\bibitem[2001]{Lehmann01} Lehmann I., Hasinger G., Schmidt M., et al., 2001, A\&A, 371,833
\bibitem[1986]{Lockman86} Lockman F.J., Jahoda K. \& McCammon D., 1986, ApJ, 302, 432
\bibitem[1991]{Maccacaro91} Maccacaro T., Della Ceca R., Gioia I.M., et al., 1991, ApJ, 374, 117
\bibitem[1998]{Magorrian98} Magorrian J., Tremaine S., Richstone D., et al., 1998, AJ, 115, 2285
\bibitem[2004]{Marconi04} Marconi A., Risaliti G., Gilli R., et al., 2004, MNRAS, 351, 169
\bibitem[1983]{Marshall83} Marshall H.L., Avni Y., Tananbaum H. \& Zamorani G., 1983, ApJ, 269, 35
\bibitem[2000]{Mason00} Mason K.O., Carrera F.J., Hasinger G., et al., 2000, MNRAS, 311, 456
\bibitem[2005]{Mateos05} Mateos S., Barcons X., Carrera F.J., et al., 2005, A\&A, 433, 855
\bibitem[2004]{Merloni04} Merloni A., 2004, MNRAS, 353, 1035
\bibitem[2000]{Miyaji00} Miyaji T., Hasinger G. \& Schmidt M., 2000, A\&A, 353, 25
\bibitem[2001]{Miyaji01} Miyaji T., Hasinger G. \& Schmidt M., 2001, A\&A, 369, 49
\bibitem[2000]{Page00} Page M.J. \& Carrera F.J., 2000, MNRAS, 311, 433
\bibitem[1997]{Page97} Page M.J., Mason K.O., McHardy I.M., et al., 1997, MNRAS, 291, 324
\bibitem[2004]{Perola04} Perola G.C., Puccetti S., Fiore F., et al., 2004, A\&A, 421, 491
\bibitem[1998]{Richstone98} Richstone D., Ahjar E.A., Bender R., et al., 1998, Nature, 395, 14
\bibitem[2002]{Rosati02} Rosati P., Tozzi P., Giacconi R., et al., 2002, ApJ, 566, 667
\bibitem[2007]{Sazonov07} Sazonov S., Revnivtsev M., Kryvonos R., et al., 2007, A\&A, 462, 57
\bibitem[1968]{Schmidt68} Schmidt M., 1968, ApJ, 151, 393
\bibitem[2000]{Schwope00} Schwope A.D., Hasinger G., Lehmann I., et al., 2000, AN, 321, 1
\bibitem[1989]{Setti89} Setti G. \& Woltjer L., 1989, A\&A, 224, L21
\bibitem[2008]{Silverman08} Silverman J.D., Green P.J., Barkhouse W.A., et al., 2008, ApJ, 679, 118
\bibitem[2005]{Simpson05} Simpson C., 2005, MNRAS, 360, 565
\bibitem[1982]{Soltan82} Soltan A., 1982, MNRAS, 200, 115
\bibitem[2003]{Spergel03} Spergel D.N., Verde L., Peiris H.V., et al., 2003, ApJS, 148, 175
\bibitem[2003]{Steffen03} Steffen A.T., Barger A.J., Cowie L.L., et al., 2003, ApJ, 569, L23
\bibitem[2004]{Szokoly04} Szokoly G.P., Bergeron J., Hasinger G., et al., 2004, ApJS, 155, 271
\bibitem[2006]{Tozzi06} Tozzi P, Gilli R., Mainieri V., Norman C., et al., 2006, A\&A, 451, 457
\bibitem[2006]{Treister06} Treister E. \& Urry C.M., 2006, ApJL, 652, 79
\bibitem[2003]{Ueda03} Ueda Y., Akiyama M., Ohta K. \& Miyaji T., 2003, ApJ, 598, 886
\bibitem[2005]{Ueda05} Ueda Y., Ishisaki Y., Takahashi T., et al., 2005, ApJS, 161, 185
\bibitem[1999]{Voges99} Voges W., Aschenbach B., Boller T., et al., 1999, A\&A, 349, 389
\bibitem[2001]{Wolf01} Wolf C., Dye S., Kleinheinrich M., et al., 2001, A\&A, 377, 442
\bibitem[2003]{Wolf03} Wolf C., Meisenheimer K., Rix H.W., et al., 2003, A\&A, 401, 73
\bibitem[2004]{Wolf04} Wolf C., Meisenheimer K., Kleinheinrich M., et al., 2004, A\&A, 421, 913
\bibitem[2002]{Yu02} Yu Q. \& Tremaine S., 2002, MNRAS, 335, 965
\bibitem[2004]{Zheng04} Zheng W., Mikles V.J., Mainieri V., et al., 2004, ApJS, 155, 73


\end{thebibliography}
\end{document}